\documentclass[prd,nofootinbib,floats,superscriptaddress,eqsecnum,tightenlines,preprintnumbers,11pt]{revtex4-1}

\usepackage{hyperref}
\usepackage{graphicx}
\usepackage{amsmath,amssymb,amsfonts,amsthm,latexsym,stmaryrd}
\usepackage{marginnote}
\usepackage{xcolor}

\usepackage{physics}
\usepackage{csquotes}
\usepackage{tensor}
\usepackage{mathtools}
\usepackage{amssymb}
\usepackage{siunitx}
\usepackage{caption}
\usepackage{cleveref}
\usepackage{multirow}
\usepackage{subcaption}
\usepackage{rotating}
\usepackage[toc,page]{appendix}
\usepackage{nicefrac}

\newcommand{\gb}{\mathcal{G}}

\newcommand{\bea}{\begin{eqnarray}}
	\newcommand{\eea}{\end{eqnarray}}
\newcommand{\beq}{\begin{equation}}
	\newcommand{\eeq}{\end{equation}}

\newcommand\cB{\Phi} 
\newcommand\cC{\Gamma} 
\newcommand\cD{\Delta} 

\newcommand\Cm{C}  
\newcommand\Cg{\hat{C}}  
\newcommand\hc{h_c}  

\newcommand\X{Y}

\begin{document}
	
\title{Linear perturbations of Einstein-Gauss-Bonnet black holes}

\author{David Langlois}
\affiliation{Universit\'e Paris Cit\'e,  CNRS, Astroparticule et Cosmologie, F-75013 Paris, France}

\author{Karim Noui}
\affiliation{Universit\'e Paris-Saclay, CNRS/IN2P3, IJCLab, 91405 Orsay, France}
\affiliation{Universit\'e Paris Cit\'e,  CNRS, Astroparticule et Cosmologie, F-75013 Paris, France}

\author{Hugo Roussille}
\affiliation{Universit\'e Paris Cit\'e,  CNRS, Astroparticule et Cosmologie, F-75013 Paris, France}
\affiliation{Universit\'e Paris-Saclay, CNRS/IN2P3, IJCLab, 91405 Orsay, France}

\date{\today}

\begin{abstract}
	We study linear perturbations about non rotating black hole solutions in scalar-tensor theories, more specifically  Horndeski theories. We consider two particular theories  that  admit known hairy black hole solutions. The first one,  Einstein-scalar-Gauss-Bonnet theory,  contains a Gauss-Bonnet term coupled to a scalar field, and its black hole solution is given as a  perturbative expansion in a small parameter that measures the deviation from general relativity. The second one, known as 4-dimensional-Einstein-Gauss-Bonnet theory,   can be seen as a compactification of higher-dimensional Lovelock theories and admits an exact black hole solution. We study both axial and polar perturbations about these solutions and write their equations of motion as a first-order (radial) system of differential equations, which enables us to study the asymptotic behaviours of the perturbations at infinity and at the horizon following an algorithm we developed recently. For the axial  perturbations,  we also obtain  effective Schr\"odinger-like equations with explicit expressions for the potentials and the propagation speeds. We see that while the Einstein-scalar-Gauss-Bonnet solution has well-behaved perturbations, the solution of the 4-dimensional-Einstein-Gauss-Bonnet theory exhibits unusual asymptotic behaviour of its perturbations near its horizon and at infinity, which makes the definition of ingoing and outgoing modes impossible. This indicates that the dynamics of these perturbations  strongly differs from the general relativity case and seems pathological.
\end{abstract}

\maketitle

\section{Introduction}
With the advent of gravitational wave (GW) astronomy, it is now possible to explore directly, via GW signals,  the strong gravity regime that characterises the merger of two black holes. So far, GW observational data are in agreement with the general relativity (GR) predictions, but it is important to test this further with more precise and more abundant data that will be available in the near future. In parallel, as a way to guide the analysis of future data, it is useful to anticipate possible deviations from the GR predictions by exploring alternative theories of gravity.  While the full description of a black hole merger in a model of modified gravity might be a daunting task, given the complexity that it already represents in  GR, the ringdown phase of the merger appears simpler to consider in a broader range of theories, since it involves the study of perturbations of a single black hole. It is nevertheless already a challenging task since black hole solutions in modified gravity theories are much more involved than GR solutions.

In the present work, we  restrict our study to the case of non rotating black holes in scalar tensor theories, corresponding to the case of most known solutions. The most general  family of scalar-tensor theories with a single scalar degree of freedom are known as degenerate higher-order scalar-tensor (DHOST) theories  \cite{Langlois:2015cwa,Langlois:2015skt,Crisostomi:2016czh,BenAchour:2016fzp}
and 
perturbations of non rotating black holes within  this family or sub-families, such as Horndeski theories \cite{Horndeski:1974wa}, have been investigated in several works.
For black hole solutions in  Horndeski theories with a purely radially dependent scalar field,  the axial perturbations  were investigated in \cite{Kobayashi:2012kh} and the polar perturbations  in \cite{Kobayashi:2014wsa},  in both cases
by  reducing the quadratic action for linear perturbations to extract  the physical degrees of freedom. This analysis was extended in \cite{Ogawa:2015pea,Takahashi:2016dnv} to include a linear time dependence of the background scalar profile, although the stability issue  was subsequently revisited  in \cite{Babichev:2018uiw}.  Axial modes were further discussed in \cite{Takahashi:2019oxz} and   \cite{Tomikawa:2021pca}  in the context of general DHOST theories.
The perturbations of stealth black holes in  DHOST theories  were investigated  in  \cite{deRham:2019gha,Khoury:2020aya} and more recently in  \cite{Takahashi:2021bml}. Note that, beyond non-rotating black holes, perturbations of the stealth Kerr black hole solutions found in \cite{Charmousis:2019vnf}  were analysed in \cite{Charmousis:2019fre}.

The approach adopted in \cite{Kobayashi:2012kh,Kobayashi:2014wsa,Takahashi:2021bml}  relies on the definition of master variables in order to rewrite the quadratic Lagrangian for perturbations in terms of the physical degrees of freedom. The procedure to identify the master variables can however be quite involved, as illustrated in \cite{Takahashi:2021bml} for stealth black holes. It is moreover strongly background-dependant and a general procedure might not exist. In \cite{Langlois:2021xzq} and \cite{Langlois:2021aji}, we  introduced another approach which focuses on the asymptotic behaviours of the perturbations, allowing to identify the physical degrees of freedom in the asymptotic regions, namely at spatial infinity and near the horizon. Since quasinormal modes are defined by specific boundary conditions (outgoing at spatial infinity and ingoing at the horizon), this is in principle sufficient to understand their properties and compute them numerically. Their asymptotic behaviour can also be used as a starting point to solve numerically the first-order radial equations and thus obtain the QNMs complex frequencies and the corresponding radial profiles of the modes.

In the present paper, we study scalar-tensor theories involving a Gauss-Bonnet term and we focus our attention on  two types of models. First, we consider Einstein-scalar-Gauss-Bonnet (EsGB) theories, which contain a scalar field with a standard kinetic term and  is coupled to the Gauss-Bonnet combination. 
We also investigate a specific scalar-tensor theory (4dEGB) that can be seen as a  4d limit of Gauss-Bonnet, in which it is possible to find an exact black hole solution \cite{Lu:2020iav} (see also \cite{Hennigar:2020lsl}).
Both EsGB and 4dEGB  models belong to DHOST theories, and more specifically to the Horndeski  theories (which we prove using the expression of Lovelock invariants as total derivatives given in \cite{colleauxRegularBlackHole2019}). They however involve cubic terms in second derivatives of the scalar field. We thus need to slightly extend the formalism introduced in \cite{Langlois:2021aji}, which was limited to terms up to quadratic order,  to include these additional terms.

Perturbations of non rotating black holes in  EsGB theories have been investigated numerically in \cite{Blazquez-Salcedo:2016enn} and \cite{Blazquez-Salcedo:2017txk}. In the present work, we revisit the analysis of these perturbations by applying  our asymptotic formalism to the perturbative description of EsGB black holes recently presented in \cite{Julie:2019sab}. We also give the Schr\"odinger-like equation for the axial modes.

Concerning the 4dEGB black hole, the present work is,  to our knowledge, the first investigation of its perturbations. For axial perturbations, the system contains a single degree of freedom and one can reformulate the equations as a Schr\"odinger-like equation. 

The outline of the paper is the following. In the next section, we introduce and extend the formalism that describes linear perturbations about static spherically symmetric solutions in scalar-tensor theories, allowing for second derivatives of the scalar field up to cubic order in the Lagrangian. Section \ref{section_EsGB} focusses on Einstein-scalar-Gauss-Bonnet theories. After discussing the background solution, which is known analytically only in a perturbative expansion, we consider first the axial modes and then the polar modes. We write their equations of motion and find their
asymptotic behaviours near the horizon and at infinity, which is a necessary requirement to define and compute quasi-normal modes. We then turn, in section \ref{section_4dGB},   to the 4dEGB black hole solution which is treated similarly. We can find that while the Einstein-scalar-Gauss-Bonnet solution has well-behaved perturbations, the solution of the 4-dimensional-Einstein-Gauss-Bonnet theory exhibits unusual asymptotic behaviour of its perturbations near its horizon and at infinity, which makes the definition of ingoing and outgoing modes impossible. We  discuss these results and conclude with a summary and some perspectives. Technical details are given in several appendices.

\section{First-order system for Horndeski theories}

In this work,  we study models that belong to  Horndeski theories,  which are included in the general family of DHOST theories.
Their  Lagrangian can be written in the form
\begin{multline}
	S[g_{\mu\nu}] = \int \dd[4]{x} \sqrt{-g} \Big(F {}^{(4)}\! R + P + Q \square\phi + 2 F_X(\phi_{\mu\nu}\phi^{\mu\nu} - \square\phi^2) + G E_{\mu\nu} \phi^{\mu\nu} \\ + \frac13 G_X (\square\phi^3 -3\square\phi \phi_{\mu\nu}\phi^{\mu\nu} + 2 \phi_{\mu\rho} \phi^{\rho\nu} \tensor{\phi}{_\nu^\mu})\Big) \,,
	\label{eq:generic-quintinc-Horndeski}
\end{multline}
where ${}^{(4)}\! R$ is  the Ricci scalar for the metric $g_{\mu\nu}$, {$E_{\mu\nu}$ is the Einstein tensor},  and we use the short-hand notations $\phi_\mu \equiv \nabla_\mu \phi$ and $\phi_{\mu\nu} \equiv \nabla_\nu \nabla_\mu \phi$ for the first and second (covariant) derivatives of $\phi$ (we have also noted $\square\phi\equiv\phi_\mu^\mu$). The functions   $F$, $P$, $Q$ and $G$ generically depend on the scalars $\phi$ and $X\equiv \phi_\mu\phi^\mu$ and a subscript $X$ denotes a partial derivative with respect to $X$. In the following, we will consider only shift-symmetric theories, where these functions depend only on $X$.

In a theory of the above type, we consider a non-rotating black hole solution, characterised by a static and spherically symmetric metric, which can be written as 
\begin{equation}
	\label{metric}
	\dd{s}^2 = - A(r) \dd{t}^2 + \frac{1}{B(r)} \dd{r}^2 + \Cg(r) \dd{\Omega}^2 \,,\qquad  \Cg(r)\equiv \Cm(r) r^2
\end{equation}
and a scalar field of the form
\begin{equation}
	\phi(t,r)=qt +\psi(r) \, .
	\label{eq:scal-ansatz}
\end{equation}
We have included here a linear time dependence of the scalar field, which is possible for shift symmetric theories  and was discussed in particular  in  \cite{Charmousis:2021npl}  for 4dEGB black holes, but later we will assume $q=0$.
In the rest of this section, we discuss the axial and polar perturbations in general, before specialising our discussion to the specific cases of EsGB and 4dEGB black holes in the subsequent sections. 

\subsection{First-order system for axial perturbations}
\label{sectionfirstordersystemaxial}
Axial perturbations correspond to the perturbations of the metric that transform like $(-1)^{\ell}$ under parity transformation, when decomposed into spherical harmonics, where $\ell$ is the usual multipole integer. Writing the perturbed metric as \begin{equation}
	g_{\mu\nu} = \overline{g}_{\mu\nu} + h_{\mu\nu} \,,
	\label{eq:def-perts}
	\end{equation}
where $\overline{g}_{\mu\nu}$ denotes the background metric \eqref{metric} and  $h_{\mu\nu}$ the metric perturbations, and working in the traditional  Regge-Wheeler gauge and in the frequency domain,  the nonzero axial perturbations depend only on two families of functions $h_0^{\ell m}$ and $h_1^{\ell m}$, namely
\begin{eqnarray}
	&&h_{t\theta} = \frac{1}{\sin\theta}  \sum_{\ell, m} h_0^{\ell m}(r) \partial_{\varphi} {Y_{\ell m}}(\theta,\varphi) e^{-i\omega t}, \qquad
	h_{t\varphi} = - \sin\theta  \sum_{\ell, m} h_0^{\ell m}(r) \partial_{\theta} {Y_{\ell m}}(\theta,\varphi) e^{-i\omega t}, \nonumber \\
	&&h_{r\theta} =  \frac{1}{\sin\theta}  \sum_{\ell, m} h_1^{\ell m}(r)\partial_{\varphi}{Y_{\ell m}}(\theta,\varphi)e^{-i\omega t}, \qquad
	h_{r\varphi} = - \sin\theta \sum_{\ell, m} h_1^{\ell m}(r)  \partial_\theta {Y_{\ell m}}(\theta,\varphi)e^{-i\omega t}, \label{eq:odd-pert}
\end{eqnarray}
while  the scalar field  perturbation is zero by construction for axial modes. In the following, we drop the $(\ell m)$ labels  to shorten the notation.

As discussed in App.~\ref{app:axial-first-order}, the system of 10 linearised metric equations for $h_0$ and $h_1$ 
can be cast into a two-dimensional system by considering only the $(r, \theta)$ and $(\theta, \theta)$ components  of the equations and using the 2-dimensional vector
\begin{equation}
	\label{defofY}
	{}^t Y = \begin{pmatrix} h_0\,, & \hc \end{pmatrix} \, , \qquad \hc \equiv \frac{1}{\omega}\left(h_1+\Psi \,  h_0\right) \, ,
\end{equation}
where we have introduced the function
\beq
\Psi= \frac{2q}{\mathcal{F}} \left(F_X \psi' - \frac{A'}{4A} X G_X\right) \,,
\eeq
whose  denominator $\mathcal F$ has been defined by
\beq
\label{Fc}
\mathcal{F} \equiv -2q^2 F_X + A(F - 2XF_X) + \frac12 B A' \psi' X G_X\,.
\eeq
Note that $\Psi$  vanishes when $q=0$.

The resulting first-order system is of the form
\begin{equation}
	\dv{Y}{r}  = M \, Y \, , \qquad M = \begin{pmatrix}
		\frac{\Cg'}{\Cg} + i \omega \Psi& -i\omega^2 + \frac{2 i \lambda \Phi}{\Cg} \\
		- i \Gamma & \Delta + i \omega \Psi
	\end{pmatrix} \,,
	\label{eq:system-axial-canonical}
\end{equation}
where the matrix coefficients depend on the theory and on the background solution,  according to the expressions
\begin{align}
	\label{matrix_coeffs_1}
	& \Phi = \frac{\mathcal{F}}{F - 2X F_X  + \frac12 B \frac{\Cg'}{\Cg} X G_X \psi'} \,, \quad
	\Delta = - \dv{}{r}\left(\ln(\sqrt{\frac{B}{A}} \mathcal{F})\right) \,, \\
	&\Gamma = \frac{F}{B \mathcal{F}} + \frac{2q^2 F_X}{AB\mathcal{F}} + \Psi^2 + \frac{G_X X (X' - \frac{q^2 A'}{A^2})}{2 B \psi' \mathcal{F}}  \, .
	\label{matrix_coeffs_2}
\end{align}
We have also introduced
	the parameter 
	\beq
	\lambda\equiv {(\ell -1)(\ell +2)}/{2}\,,
	\eeq
	which we will be using instead of $\ell$.
Notice that the matrix M coincides with the results of \cite{Langlois:2021aji} in the cases where $G=0$ and $\Cg=r^2$.

\subsection{Schr\"odinger-like  equation and potential}
\label{subsection_schroedinger}

As discussed in detail in \cite{Langlois:2021aji}, one can introduce a new coordinate variable $r_*$ defined from a function $n(r)$ according to
\begin{eqnarray}
	\label{change_radius}
	\dv{r_*}{r} = \frac{1}{n(r)} \, , 
\end{eqnarray}
and find a linear change of functions of the form
\beq
\label{change_functions}
Y = \hat P(r) \hat \X \, , \qquad {}^t \hat \X = \begin{pmatrix} \hat\X_1, & \hat\X_2\end{pmatrix} \, ,
\eeq
such that   the initial system  \eqref{eq:system-axial-canonical}  is transformed into an equivalent system in the  ``canonical" form
\begin{equation}
	\label{eq:eqs-matrix-odd}
	\dv{\hat \X}{r_*} \equiv  n(r) \dv{\hat \X}{r}= \begin{pmatrix}  i \omega \Psi n  & 1 \\ V - {\omega^2}n^2\Gamma\quad & i \omega \Psi n \end{pmatrix}\hat \X \, .
\end{equation}
At this stage $n(r)$ is arbitrary 
and the function $V$ is given in terms of the functions characterising the theory and of the background metric by the expression 
\bea
\label{generalpotentialtext}
V  &= & \frac{n^2}{4} \left[ 8\frac{\lambda \cB \cC}{\Cg}+ \cD^2 + 2 \cD' + 
\frac{2\cC'}{\cC} \left( \frac{\Cg'}{\Cg} - \cD \right) -2 \cD\frac{\Cm'}{\Cm}  
\right. \nonumber \\
&&\left. \qquad + 3 \left( \frac{\cC'}{\cC} \right)^2 + \left(\frac{n'}{n} \right)^2+3\left(\frac{\Cg'}{\Cg}\right)^2
- 2  \left(\frac{\cC''}{\cC} + \frac{n''}{n} +\frac{\Cg''}{\Cg}\right)  
\right]\,.
\eea
This formula generalises the result given in \cite{Langlois:2021aji} to an arbitrary function $\Cg(r)$.

\medskip
When $\Psi=0$, which is the case for $q=0$, the system (\ref{eq:eqs-matrix-odd}) immediately leads to the Schr\"odinger-like second-order equation for the function $\hat\X_1$,
\beq
\label{schroedinger}
\hat\X_1''+\left(\frac{\omega^2}{c^2}-V\right)\hat\X_1=0\, ,
\eeq
which corresponds to a wave equation, written in the frequency domain, with a potential $V$ and a propagation speed given by
\beq
c\equiv  \frac{1}{n\,  \sqrt{\cC}}\,.
\label{eq:def-c}
\eeq
{As expected, the speed of propagation}  depends on $n$, i.e. on the choice of the radial coordinate.

When $\Psi\neq 0$, one can still get an equation of the form (\ref{schroedinger}), but after the change of time variable
\beq
t \longrightarrow t  - \int \Psi(r) \dd{r}\,,
\label{eq:chgvar-t-Psi}
\eeq
or equivalently the redefinition $\hat\X  \longrightarrow e^{i \omega \int \Psi(r) \dd{r}} \,\hat\X$. {Notice that such a change of variable is defined only if $\Psi$ is integrable. For instance, in the case where $\Psi$ is singular with a pole of order $1$ at some radius $r_p$,  the change of variables \eqref{eq:chgvar-t-Psi} is only valid for $r > r_p$.}

\subsection{First-order system for polar modes}

For the polar (or even-parity) perturbations we choose the same (Zerilli) gauge fixing as usually adopted in GR.  The metric perturbations are now parametrised 
by four families of functions $H_{0}^{\ell m}$, $H_{1}^{\ell m}$, $H_{2}^{\ell m}$ and $K^{\ell m}$  such that the non-vanishing components of the metric are
\bea
\label{eq:even-pert-Horn}
&&h_{tt} = A(r)\sum_{\ell, m} H_{0}^{\ell m}(t,r) Y_{\ell m}(\theta,\varphi), \quad
h_{rr} = B(r)^{-1} \sum_{\ell, m} H_{2}^{\ell m}(t,r) Y_{\ell m}(\theta,\varphi) ,  \nonumber \\
&&h_{tr} = \sum_{\ell, m} H_{1}^{\ell m}(t,r) Y_{\ell m}(\theta,\varphi), \quad h_{ab} = \sum_{\ell, m} K^{\ell m}(t,r) g_{ab} Y_{\ell m}(\theta,\varphi) \, , 
\eea
where the indices $a, b$ belong to $\{\theta, \varphi\}$. 
The scalar field perturbation is parametrised by one more family of functions $\delta\phi^{\ell m}$
according to
\begin{equation}
	\delta\phi =  \sum_{\ell, m} \delta\phi^{\ell m}(t, r) Y_{\ell m}(\theta, \varphi) \, .
\end{equation}
In the following we will consider only the modes  $\ell \geq 2$ (the monopole $\ell=0$ and the dipole $\ell=1$ require different gauge fixing conditions) {and we drop $(\ell m)$ labels to lighten notations}.

One can show that the $(t, r)$, $(r, r)$, $(t, \theta)$ and $(r, \theta)$ components of the perturbed metric equations, which are first order in radial derivatives, are sufficient to describe the dynamics of the perturbations \cite{Langlois:2021aji}. Therefore, the linear equations of motion can be written as a first-order differential system,
\bea
\frac{\mathrm{d}\X}{\mathrm{d}r}=M \X \,,
\eea
satisfied by the four-dimensional  vector 
\begin{equation}
	\X = {}^t\!\begin{pmatrix}
		K, & \chi, & H_1, & H_0
	\end{pmatrix} \,,
\end{equation}
where $\chi$ is proportional to the scalar field perturbation. The precise proportionality factor depends on the background solution and will be given   explicitly later  in each of the two cases considered in this paper. The form of the  square matrix $M$ can be read off from the equations of motion.

\section{Einstein-scalar-Gauss-Bonnet black hole}
\label{section_EsGB}
In this section, we specialise our study to the case of Einstein-scalar-Gauss-Bonnet (EsGB) theories, where one adds to the usual Einstein-Hilbert term for the metric, a non-standard coupling to a scalar field $\phi$ which involves the Gauss-Bonnet term \eqref{eq:action-phi-GB}. Analytical non rotating black hole solutions were found in the case of specific coupling values in \cite{Mignemi:1992nt,Torii:1996yi,Yunes:2011we,Sotiriou:2013qea,Sotiriou:2014pfa}, and rotating solutions in the same setups in \cite{Ayzenberg:2014aka,Pani:2011gy,Maselli:2015tta}. A solution for any coupling form was obtained in \cite{Julie:2019sab}, and a solution with an additional cubic galileon coupling was proposed in \cite{Hui:2021cpm}. All these solutions are given as expansions in a small parameter appearing in the coupling function. This small parameter parametrises the deviation from GR.

\subsection{Action}
The Einstein-scalar-Gauss-Bonnet  action is given by
\begin{equation}
	S[g_{\mu\nu}] = \int \dd[4]{x} \sqrt{-g} \left(R - 2 X +  f(\phi) \gb \right) \,,
	\label{eq:action-phi-GB}
\end{equation}
where   $f(\phi)$ is an arbitrary function of $\phi$  and
\begin{equation}
	\label{GBinvariant}
	\gb = R_{\mu\nu\rho\sigma} R^{\mu\nu\rho\sigma} - 4 R_{\mu\nu} R^{\mu\nu} + R^2 \,,
\end{equation}
is the Gauss-Bonnet term in 4 dimensions.

Although this action is not manifestly of  the form (\ref{eq:generic-quintinc-Horndeski}), its equations of motion can be shown to be second order, which means that the theory  can be reformulated as a Horndeski theory \cite{Kobayashi:2019hrl, Kobayashi:2011nu}. This is explicitly shown in Appendix \ref{app:equivalence-EGB-Horndeski}, working directly at the level of the action. The corresponding Horndeski functions are given by
\begin{align}
	&P(\phi, X) = -2 X + 2 {f^{(4)}}(\phi) \, X^2 (3 - \ln X) \,,\quad Q(\phi, X) = 2  {f^{(3)}}(\phi) \, X (7 - 3 \ln X) \,\nonumber\\
	&F(\phi, X) = 1 - 2 {f''}(\phi) \, X (2 - \ln X) \qq{and} G(\phi, X) = -4   f'(\phi)\,  \ln X \,.
\end{align}
Here, we are using the notation ${f^{(n)}}(\phi)$ for the $n$-th derivative of $f(\phi)$ with respect to $\phi$.

\subsection{Background solution}

To find a static black hole solution, we start with the ansatz
\begin{equation}
	\dd{s}^2 = -A(r) \dd{t}^2 + \frac{1}{A(r)} \dd{r}^2 + r^2 \Cm(r) \dd{\Omega}^2 \qq{and} \phi = \psi(r)\,,
\end{equation}
corresponding to the gauge choice $B=A$ and $\Cm\neq 1$ in (\ref{metric}). An alternative choice would have been to assume $C=1$ and $B\neq A$  (see for example \cite{Bryant:2021xdh}).

When the coupling function $f$ is a constant,  the { term proportional to $\cal G$ in the action} becomes a total derivative and is thus irrelevant for the equations of motion, which are then the same as in GR with a massless scalar field. One thus  immediately obtains as a  solution the Schwarzschild metric with a constant and uniform scalar field,   
\begin{equation}
	A(r) = 1 - \frac{\mu}{r} \,, \quad C(r) = 1 \qq{and} \psi(r) = \psi_\infty \,,
	\label{eq:background-EGB-epsilon}
\end{equation}
where $\psi_\infty$ is an arbitrary constant. 

When $f(\phi)$ is not constant, the above configuration is no longer a solution  but can nevertheless be considered as the zeroth order 
expression of the full solution  written as  a series expansion  in terms  of the parameter 
\begin{equation}
	\varepsilon = \frac{f'(\psi_\infty)}{\mu^2}  \,,
\end{equation}
assumed to be small, as  it was proposed initially proposed in \cite{Mignemi:1992nt} and recently developed in \cite{Julie:2019sab}. Hence, we expand  the metric components and scalar field  as series in power of $\varepsilon$  (up to some order $N$) as follows,
\begin{align}
	\label{A_epsilon}
	A(r) &= 1 - \frac{\mu}{r} + \sum_{i = 1}^N a_i(r) \varepsilon^i + \mathcal{O}(\varepsilon^{N+1}) \,,\\
	C(r) &= 1 + \sum_{i = 1}^N c_i(r) \varepsilon^i + \mathcal{O}(\varepsilon^{N+1}) \,,\\
	\psi(r) &= \psi_\infty + \sum_{i = 1}^N s_i(r) \varepsilon^i + \mathcal{O}(\varepsilon^{N+1}) \,,
\end{align}
where  the functions $a_i$, $c_i$ and $s_i$ can be determined, order by order, by solving the associated differential equations obtained by substituting the above expressions into   the equations of motion. 

One can see that the metric equations of motion expanded up to order $\varepsilon^N$ involve $a_N(r)$, $c_N(r)$ and $s_{N-1}'(r)$, while the scalar equation of motion at order $\varepsilon^N$ relates $a_{N-1}(r)$, $c_{N-1}(r)$ and $s_N'(r)$. Then, it is possible to use this separation of orders to solve the equations of motion order by order. We need boundary conditions to integrate these equations and we impose that all these functions go to zero at spatial infinity.

\medskip

At  first order in $\varepsilon$, one obtains the equations
\begin{align}
	a_1(r) = - \tau_3 + \frac{1}{r} \qty(\tau_1+ \tau_2) - \frac{\mu\tau_2}{2r^2} \,,\qquad 
	c_1(r) = \tau_3 - \frac{\tau_2}{r} \,,
\end{align}
where the $\tau_i$ are integration constants. The boundary conditions at spatial infinity impose $\tau_3 = 0$. Furthermore, the constant $\tau_1 + \tau_2$, which  can be interpreted as a shift of the black hole mass at first order in $\varepsilon$, can be absorbed by redefining $\mu$  as follows:
\begin{equation}
	\mu_{\rm new} = \mu_{\rm old} - \varepsilon(\tau_1 + \tau_2) \,.
\end{equation}
Finally, the remaining terms proportional to $\tau_2$ can be absorbed by the coordinate change  
\begin{eqnarray}
	r_{\rm new}= r_{\rm old} +\varepsilon \tau_2/2 \, . 
\end{eqnarray}
As a consequence, at first order in $\varepsilon$, one  simply recovers the background solution given in Eq.~\eqref{eq:background-EGB-epsilon}, up to a change of mass and a change of coordinate, which corresponds to taking
\begin{equation}
	a_1(r) = 0 \qq{and} c_1(r) = 0 \,.
\end{equation}
As for the scalar field, its equation of motion yields, at first order in $\varepsilon$,
\begin{equation}
	s_1(r) = \frac{\mu}{r} + \frac{\mu^2}{2r} + \frac{\mu^3}{3r^3} + \nu_1 + \qty(1+ \frac{\nu_2}{\mu}) \ln(1 - \frac{\mu}{r}) \,,
\end{equation}
with $\nu_1$ and $\nu_2$ constants. One can obviously absorb the constant $\nu_1$ into a redefinition of $\psi_\infty$ while one chooses  
$\nu_2$ so that $s_1(r)$ remains regular  at the horizon. 

At order  $\varepsilon^2$, one can repeat the same method to solve for $a_2$, $b_2$ and $s_2$.  One can ignore the five integration constants that appear  since they can be reabsorbed  using the boundary conditions, mass redefinition and coordinate change, as previously. At the end, the metric and scalar functions  read
\begin{align}
	a_2(r) &= - \qty(\frac{\mu^3}{3r^3} - \frac{11\mu^4}{6r^4} + \frac{\mu^5}{30r^5} + \frac{17\mu^7}{15r^7})\,, \label{a_2}\\
	c_2(r) &= - \qty(\frac{\mu^2}{r^2} + \frac{2\mu^3}{3r^3} + \frac{7\mu^4}{6r^4} + \frac{4\mu^5}{5r^5} + \frac{3\mu^6}{5r^6})\,,\\
	s_2(r) &= \rho_2 \qty(\frac{73}{60} \qty(\frac{\mu}{r} + \frac{\mu^2}{2r^2} + \frac{\mu^3}{3r^3} + \frac{\mu^4}{4r^4} ) + \frac{7\mu^5}{75r^5} + \frac{\mu^6}{36r^6})\,,
\end{align}
where we have introduced the constant $\rho_2$ defined by
\begin{equation}
	\label{defofrho2}
	\rho_2 = \frac{f''(\psi_\infty)}{f'(\psi_\infty)} \,.
\end{equation}

In principle,  it is possible to continue this procedure and find all coefficients up to some arbitrary order $\varepsilon^N$ in a finite number of steps, but the complexity of the expressions  quickly makes the computations very cumbersome. Here, we stop at order $\varepsilon^2$, but one could proceed similarly for the next orders, for instance to obtain the numerical precision in the  computation of quasinormal modes reached in \cite{Blazquez-Salcedo:2016enn}.

By taking into account  the higher order corrections to the metric functions, the  black hole horizon is no longer at $r=\mu$ but is slightly
shifted to the   new value 
\begin{equation}
	r_h = \mu\qty(1 - \frac{\varepsilon^2}{3}) + \mathcal{O}(\varepsilon^3) \,.
\end{equation}
Since  $r_h$ is known only as a power series of $\varepsilon$, it is more  convenient to work  with the new radial coordinate dimensionless 
variable $z$, 
\beq
\label{z}
z=\frac{r}{r_h}\,,
\eeq
in terms of which the horizon is exactly located at $z=1$, at any order in $\varepsilon$.

\subsection{Axial modes: Schr\"odinger-like equation}
Let us now turn to the study of perturbations about this black hole solution, starting with axial perturbations (note that perturbations with the specific choice of coupling $f(\phi) = \phi$ were studied in the context of a stability analysis in \cite{Minamitsuji:2022mlv}).
As we have seen in section \ref{sectionfirstordersystemaxial}, the first-order system for the axial modes can be written in the form \eqref{eq:system-axial-canonical} and depends only on the functions $\Gamma$, $\Delta$ and $\Phi$, since here $\Psi=0$ (because $q=0$).
In terms of the new radial coordinate $z$, these functions read, up to order $\varepsilon^2$,
\begin{align}
	\Gamma &= \frac{1}{(z-1)^2} \left[ {z^2}+\frac{10 z^5+10 z^4-100 z^3-95 z^2-94 z+206}{15  z^4}  \varepsilon^2 \right]+ \mathcal{O}(\varepsilon^3) \,,\nonumber\\
	\Phi &= (z-1) \left[ \frac{1}{z}- \frac{10 z^5 + 10 z^4 + 140 z^3 - 95 z^2 - 94 z - 214}{30 z^7}  \varepsilon^2 \right]+ \mathcal{O}(\varepsilon^3) \,,
	\label{eq:def-functions-axial-aEGB} \\
	\Delta &=\frac{1}{z-z^2}+\frac{-5 z^5-10 z^4-30 z^3+190 z^2+235 z+282}{15 z^7} \varepsilon^2 + \mathcal{O}(\varepsilon^3) \,.\nonumber 
\end{align}
When $\varepsilon$ goes to zero, one recovers  the standard Schwarzschild expressions, as computed in \cite{Langlois:2021aji}.

\begin{figure}[!htb]
	\begin{subfigure}{0.45\textwidth}
		\centering
		\includegraphics{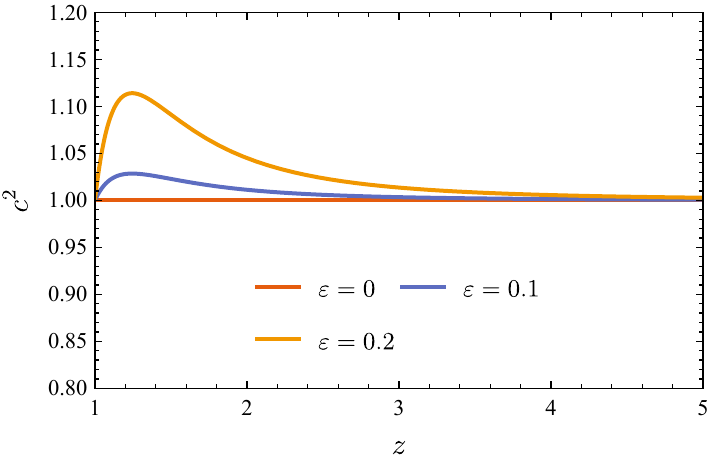}
		\caption{Squared speed}
	\end{subfigure}
	\begin{subfigure}{0.45\textwidth}
		\centering
		\includegraphics{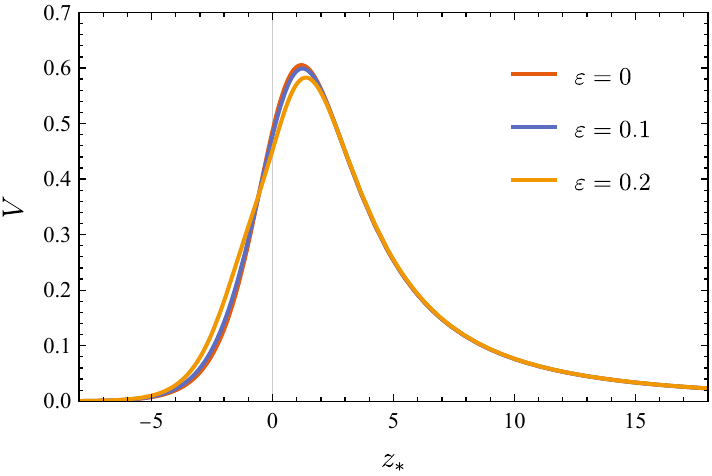}
		\caption{Potential}
	\end{subfigure}
	\caption{\small{Plot of the squared propagation speed $c^2$ as a function of $z$ and the potential $V$  as a function of $z_*$ for $\lambda=2$. Note that the coordinate $z_*$ is defined up to a constant which,  in this plot,  differs from the choice in \eqref{zstartortue}. In the figure, the constant is chosen such that $z_* = 0$ when $z = 1 + W(e^{-1})$, where  $W$ is the Lambert function.  This corresponds to the definition $z_* = z + \ln(z-1)$  in the GR case ($\varepsilon = 0$).}}
	\label{fig:c-pot-aEGB}
\end{figure}

By substituting the above expressions into \eqref{eq:def-c} and \eqref{generalpotentialtext} and choosing $n(z)=A(z)$, one can then obtain (up to order  $\varepsilon^2$)  the propagation speed from, 
\begin{align}
	c^2 &= 1 + 4 \varepsilon^2 \qty(-\frac{4}{z^6} + \frac{1}{z^5} + \frac{1}{z^4} + \frac{2}{z^3}) + \mathcal{O}(\varepsilon^3) \, ,
\end{align}
and also the potential, 
\begin{align}	
	V &= \qty(1-\frac{1}{z}) \left[ \frac{-3 + 2z(1+\lambda)}{z^3} + \varepsilon^2\Big(
	\frac{2542}{5} \frac{1}{z^9}
	+ \frac{1}{15} (-8009 + 712 \lambda) \frac{1}{z^8}
	- \frac{2}{15} (-29+\lambda) \frac{1}{z^7} \right. \nonumber\\
	&\left. \quad\quad+\frac23 (-47+ \lambda)\frac{1}{z^6}
	+ (70-24\lambda) \frac{1}{z^5}
	+\frac43(4+\lambda) \frac{1}{z^4}
	-\frac13(5+2\lambda) \frac{1}{z^3}
	\Big) \right] + \mathcal{O}(\varepsilon^3) \,.
	\label{eq:pot-speed-aEGB}
\end{align}
These quantities have been illustrated in Fig.\eqref{fig:c-pot-aEGB} for some values of $\varepsilon$.   Note that the potential is plotted  as a function of the ``tortoise" coordinate $z_*$, defined similarly to $r_*$ in \eqref{change_radius} with $n=A$:
\begin{equation}
	\dv{z_*}{z} = \frac{1}{n(z)} = \frac{1}{A(z)} \,.
	\label{eq:def-z-tortoise}
\end{equation}
Substituting  the expression of $A(z)$, obtained from \eqref{A_epsilon}, \eqref{a_2} and \eqref{z}, 
\begin{equation}
	A(z) = \qty(1 - \frac{1}{z}) - \varepsilon^2 \qty(\frac{17}{15z^7} + \frac{1}{30 z^5} - \frac{11}{6z^4} + \frac1{3z^3} + \frac1{3z}) + \mathcal{O}(\varepsilon^3) \,,
\end{equation}
one gets
\begin{equation}
	z_* = z - \varepsilon^2\qty(\frac{17}{60 z^4} + \frac{34}{45 z^3} + \frac{103}{60 z^2} + \frac{83}{30 z} - \frac{73}{30} \ln(z) ) + \ln(z- 1) \qty(1 - \frac{21}{10} \varepsilon^2) + \mathcal{O}(\varepsilon^3) \,.
	\label{zstartortue}
\end{equation}
This implies, in particular,  the asymptotic behaviours at spatial infinity
\begin{equation}
	\label{z_*_infinity}
	z_* \simeq z + \ln(z)\qty(1 + \frac{\varepsilon^2}{3}) + \mathcal{O}(\varepsilon^3) \,,  \qquad (z\to +\infty)
\end{equation}
and  at the horizon, 
\begin{equation}
	z_* \simeq \ln(z-1) \qty(1 - \frac{21}{10} \varepsilon^2)  + \mathcal{O}(\varepsilon^3) \, . \qquad (z\to 1)
	\label{eq:z-tortoise-horizon}
\end{equation}
Notice that all along the paper, we will be using the symbol $\simeq$ for  an equality up to sub-dominant terms in the $z$ variable when $z\gg 1$ at infinity and $z-1 \ll 1$ at the horizon. More precisely, given two functions $f(z)$ and $g(z)$, we say that $f(z) \simeq g(z)$ at $z_0$ (which can be here $z_0=\infty$ or $z_0=1$) when
\begin{eqnarray}
	f(z) \simeq g(z)  \quad \text{at} \; z\to z_0 \quad \text{means} \quad \lim_{z \to z_0} \frac{f(z)-g(z)}{f(z)} = 0 \, .  
\end{eqnarray}

Noting that $c$ tends to $1$ and $V$ vanishes  both at the horizon and at spatial infinity,  the asymptotic behaviour of  \eqref{schroedinger} is simply
given by %
\begin{equation}
	\dv[2]{\hat{\X}_1}{z_*} + \Omega^2 \hat{\X}_1 \, \simeq \, 0 \, ,
	\label{eq:schrodinger-asymp}
\end{equation}
where we have  rescaled  the frequency according to
\begin{equation}
	\Omega = {\omega}\, {r_h} \,.
\end{equation}
As a consequence, at spatial infinity, using  \eqref{z_*_infinity},   the asymptotic  solution is
\begin{equation}
	\hat{\X}_1 \simeq \mathcal{A}_{\infty} e^{+ i\Omega z} z^{i \Omega\qty(1 + {\varepsilon^2}/{3})} + \mathcal{B}_{\infty} e^{- i\Omega z} z^{-i \Omega\qty(1 + {\varepsilon^2}/{3})} \,,
	\label{eq:asymp-behav-infty-aEGB-schrodinger}
\end{equation}
while  the solution near  the horizon takes the form,
\begin{equation}
	\hat{\X}_1 \simeq \mathcal{A}_\text{hor} (z-1)^{+i\Omega \left(1 - {21} \varepsilon^2/10 \right)} + \mathcal{B}_\text{hor} (z-1)^{-i\Omega \left(1 - {21} \varepsilon^2/10 \right)} \,,
	\label{eq:asymp-behav-hor-aEGB-schrodinger}
\end{equation}
where we have used \eqref{eq:z-tortoise-horizon} when we replace $z_*$ by its expression in terms of $z$. Finally, the constants 
$\mathcal{A}_{\infty} $, $\mathcal{B}_{\infty} $, $\mathcal{A}_\text{hor} $ and $\mathcal{B}_\text{hor} $ can be fixed or partially fixed 
by appropriate boundary conditions.

\subsection{Axial modes: first order system and asymptotics}

In this subsection, we show that  the asymptotic solutions, obtained previously in \eqref{eq:asymp-behav-infty-aEGB-schrodinger} and \eqref{eq:asymp-behav-hor-aEGB-schrodinger} from the Schr\"odinger-like equation, can be recovered directly from  the first-order system which corresponds to \eqref{eq:system-axial-canonical} with the definitions \eqref{eq:def-functions-axial-aEGB}.  We will be making use of the algorithm presented in \cite{Langlois:2021xzq}.

\subsubsection{First order system and asymptotics: brief review and notations}
\label{subsectionalgo}
The goal of the algorithm is to find a set of functions so that the original system is reexpressed in the simpler form
\beq
\label{diagonalM}
\dv{\tilde{\X}}{x} = \tilde{M} \tilde{\X} \,, \qquad \tilde{M} = x^p \sum_{i=0}^N D_i x^i + \mathcal{O}(x^{N+1}) \,,
\eeq
where the matrices $D_i$ are diagonal\footnote{In some specific cases which are described in \cite{Langlois:2021xzq}, one can only reduce the system up to $p = -1$ without a diagonal leading order. However, this will not be the case for the system studied in this paper.} and $x$ is a new variable, defined such that the asymptotic limit considered corresponds to $x \rightarrow +\infty$. For spatial infinity, we simply use $x=z$, whereas we choose $x = 1/(z-1)$ for the near horizon limit $z\to 1$. 

More precisely, we start with the system 
\begin{equation}
	\dv{\X}{x} = M \X \, ,
\end{equation}
whose matrix $M$ is immediately obtained from \eqref{eq:system-axial-canonical} with \eqref{eq:def-functions-axial-aEGB}, then we make the change of  variable from $z$  to $x$ if necessary, and finally the algorithm described in
\cite{Langlois:2021xzq} provides us with the transfer matrix $\tilde{P}$ defining the appropriate change of functions
\begin{equation}
	\label{P_tilde}
	\X = \tilde{P}\,  \tilde{\X} \,,
\end{equation}
so that the new matrix $\tilde{M}$, given by
\beq
\label{M_tilde}
\tilde{M}= \tilde{P}^{-1} M  \tilde{P}- \tilde{P}^{-1}\, \dv{\tilde{P}}{x}\,,
\eeq
{takes the diagonal form \eqref{diagonalM}}. Hence, we obtain immediately the asympotic behaviour of the solution by integrating the diagonal first-order differential system \eqref{diagonalM}. 
We apply this procedure in turn to the spatial infinity and near horizon limits, up to order $\varepsilon^2$.

\subsubsection{Spatial infinity}
At spatial infinity, the coordinate variable is $z$ and the first terms of the initial matrix $M$ in an expansion in power of $z$ read
\begin{equation}
	M = \begin{pmatrix}
		0 & -i\Omega^2 \\ -i & 0
	\end{pmatrix} 
	+ 
	\begin{pmatrix}
		2 & 0 \\ -2i\qty(1 + {\varepsilon^2}/{3}) & 0
	\end{pmatrix} \frac{1}{z} + \mathcal{O}\qty(\frac{1}{z^2}) \,.
\end{equation}
Applying the change of functions (\ref{P_tilde}) with
\begin{equation}
	\tilde{P} = \begin{pmatrix}
		\Omega & -\Omega \\ 1 & 1
	\end{pmatrix} 
	+ \frac{1}{6\Omega}
	\begin{pmatrix}
		3i\Omega - (3+\varepsilon^2) \Omega^2 & 3i\Omega + (3+\varepsilon^2) \Omega^2 \\ -3i + (3+\varepsilon^2) \Omega & 3i + (3+\varepsilon^2) \Omega
	\end{pmatrix} \frac{1}{z} \,,
\end{equation}
provided by the algorithm of \cite{Langlois:2021xzq},  one obtains the new  matrix
\begin{equation}
	\tilde{M} = \text{diag}(-i\Omega, i\Omega) +  \frac{1}{z} \text{diag} \left[1 - i \Omega \qty(1 + \frac{\varepsilon^2}{3}),1 -+i \Omega \qty(1 + \frac{\varepsilon^2}{3}) \right] + \mathcal{O}\qty(\frac{1}{z^2}) \,,
	\label{eq:asymp-mat-axial-4EGB}
\end{equation}
which is diagonal up to order $1/z^2$. Hence,  the corresponding system can  immediately be integrated and
we  find the behaviour of $\X$ and of the metric coefficients \eqref{defofY} at infinity:
\begin{align}
	h_0(z) & \simeq  \Omega \left[-c_+ e^{i\Omega z} z^{+ i\Omega \qty(1 + {\varepsilon^2}/{3})} + c_-  e^{-i\Omega z} z^{- i\Omega \qty(1 + {\varepsilon^2}/{3})} \right]\,,\nonumber\\
	\hc(z) & \simeq c_+ e^{i\Omega z} z^{+ i\Omega \qty(1 + {\varepsilon^2}/{3})} + c_-  e^{-i\Omega z} z^{- i\Omega \qty(1 +  {\varepsilon^2}/{3})} \,, \label{axialinfinity}
\end{align}
where $c_\pm$ are the integration constants. As expected, one recovers the same combination of modes as in \eqref{eq:asymp-behav-infty-aEGB-schrodinger}.

\subsubsection{Near the horizon}

{To study the asymptotic behaviour near the horizon, it is convenient to use the coordinate $x$ defined by $x = 1/(z-1)$. Then, we study the behaviour, when $x$ goes to infinity, of the system  \eqref{eq:first-order-axial}, reformulated as
\begin{equation}
	\dv{\X}{x} = M_x(x)\,  \X\,, \qq{with}  M_x(x)= - \frac{1}{x^2} M(1 + 1/x) \,.
\end{equation}
The expansion of the matrix $M_x$ in powers of $x^{-1}$ yields}
\begin{eqnarray}
	M_x &= &\begin{pmatrix}
		0 & 0 \\ i \qty(1 - {21}/{5}\varepsilon^2) & 0
	\end{pmatrix} 
	+ 
	\begin{pmatrix}
		0 & 0 \\ 2 i \qty(1 - {121}/{15}\varepsilon^2) & 1
	\end{pmatrix}  \frac{1}{x}
	\\
	&& + \frac{1}{15}
	\begin{pmatrix}
		-30 - 244\varepsilon^2 & 15 i \Omega^2 \\ 15 i (1 + 1111 \varepsilon^2) & -15 - 662 \varepsilon^2
	\end{pmatrix}  \frac{1}{x^2}
	+ \mathcal{O}\qty(\frac{1}{x^3}) \,.
\end{eqnarray}
The algorithm provides us with the transfer matrix 
\begin{multline}
	\tilde{P} = \begin{pmatrix}
		0 & 0 \\ 1 & 1
	\end{pmatrix} x
	+ \Omega
	\begin{pmatrix}
		-1-{21}\varepsilon^2/{10} & 1 +{21}\varepsilon^2/{10}  \\ 0 & 0
	\end{pmatrix}
	\\
	+ 
	\begin{pmatrix}
		2(i + \Omega) + {\varepsilon^2} (10 i - 53 \Omega)/15 & 2(i - \Omega) + {\varepsilon^2} (10 i + 53 \Omega)/15 \\ 0 & 0
	\end{pmatrix} \frac{1}{x} 
	+ \mathcal{O}\qty(\frac{1}{x^2})
	\,,
\end{multline}
and one obtains a new differential system with a diagonal matrix $\tilde M_x$,
\begin{equation}
	\tilde{M}_x = \frac{1}{x} \text{diag} \left[ -i \Omega \qty(1 - \frac{21}{10}\varepsilon^2 ) , i \Omega \qty(1 - \frac{21}{10}\varepsilon^2 )\right]  + \mathcal{O}\qty(\frac{1}{x^2}) \,.\end{equation}
Integrating the system immediately yields the near-horizon behaviour of the metric components, expressed in terms of the variable $z$, 
\begin{align}
	h_0(z) &\simeq  \Omega \qty(1 + {21}\varepsilon^2/10 ) \left[  c_+ (z-1)^{-1 - i\Omega \qty(1 - {21}\varepsilon^2/10 )} -  c_-   (z-1)^{-1 + i\Omega \qty(1 - {21}\varepsilon^2/10 )}	\right] \,,\nonumber\\
	\hc(z) &\simeq c_+ (z-1)^{-1 - i\Omega \qty(1 - {21}\varepsilon^2/10 )} + c_- (z-1)^{-1 + i\Omega \qty(1 - {21}\varepsilon^2/10 )}\,,
\end{align}
where $c_\pm$ are integration constants (different from those introduced in \eqref{axialinfinity}). As expected,  we recover the combination of modes found in \eqref{eq:asymp-behav-hor-aEGB-schrodinger} from the Schr\"odinger-like formulation.

\subsection{Polar modes}

For polar modes, there is no obvious Schr\"odinger-like formulation of the equations of motion so the simplest approach is to work directly with the first-order system. The latter  can be written in the form
\begin{equation}
	\dv{\X}{r} = S\,  \X \,, \qq{with} \X = {}^t\!\begin{pmatrix}K, & \delta\varphi, & H_1,&  H_0 \end{pmatrix}\,,
\end{equation}
but  the matrix $S$ is singular in the GR  limit  when $\varepsilon\to 0$. This problem can be avoided by using the functions
\begin{equation}
	\label{eq:def-chi-EsGB}
	\chi = \varepsilon \, \delta\varphi \qq{and} Y = {}^t\!\begin{pmatrix}K, & \chi, & H_1, & H_0 \end{pmatrix} \,,
\end{equation}
leading to a well-defined system in the GR limit  of the form
\begin{equation}
	\dv{\X}{z} = M(z)\,  \X \, ,
	\label{eq:syst-polar}
\end{equation}
where $z$ is the dimensionless coordinate introduced in \eqref{z}. 
We now consider in turn the two asymptotic limits.

\subsubsection{Spatial infinity}
We start by computing the expansion of  $M$ in powers of $z$; we obtain
\begin{equation}
	M = M_{-2} \, z^2 + M_{-1} \, z + M_{0} + \frac{M_1}{z} + \mathcal{O}\qty(\frac{1}{z^2}) \, .
\end{equation}
The matrices $M_{-2} $, $ M_{-1}$ and $M_0 $ take the simple expressions 
\begin{eqnarray}
	&&M_{-2} = \begin{pmatrix}
		0 & 0 & 0 & 0 \\
		a & 0 & 0 & 0 \\
		0 & 0 & 0 & 0 \\
		0 & 0 & 0 & 0 
	\end{pmatrix} \, , \qquad
	M_{-1}  = \begin{pmatrix}
		0 & 0 & 0 & 0 \\
		a + \varepsilon^2\Omega^2/6 & 0 &a/i\Omega &  0 \\
		0 & 0 & 0 & 0 \\
		0 & 0 & 0 & 0 
	\end{pmatrix} \, , \\
	&& M_0  = \begin{pmatrix}
		0 &0 &0 &0 \\
		a\qty(1 - {\lambda}/{\Omega^2}) - {\varepsilon^2\Omega^2}/{3} & 0 & 0 & a {\lambda}/{\Omega^2} \\
		-i \Omega & 0 & 0 & -i\Omega \\
		0 & 0 & -i\Omega & 0
	\end{pmatrix} \, ,
\end{eqnarray}
which depend on the coefficient $a$ defined by
\begin{equation}
	a = -\frac{\Omega ^2}{2}  +\frac{73}{120} \rho_2  \Omega ^2 \varepsilon+ \frac{\Omega ^2  \left(13201 \rho_2 ^2+62555 \rho_3 +209160\right)}{151200}\, \varepsilon^2 \qq{and}
	\rho_3 = \frac{f'''(\psi_\infty)}{f'(\psi_\infty)} \,,
\end{equation}
while $\rho_2$ has been defined in \eqref{defofrho2}. The matrices $M_i$ with $i \geq 1$ are more involved than the three above matrices and we do not give their expressions here. Nonetheless, some of them enter in the algorithm briefly recalled earlier,  in section \ref{subsectionalgo}, which enables us to diagonalise the differential system \eqref{eq:syst-polar}. 

The asymptotical diagonal form at infinity cannot immediately be obtained from equation~\eqref{eq:syst-polar}, as the leading order matrix  $M_{-2} $ is nilpotent. As discussed in \cite{Langlois:2021xzq},  for this special subcase of the algorithm, one must first  obtain a {\it diagonalisable} leading order term, by applying a change of functions parametrised by the matrix 
\begin{equation}
	P^{(1)} = \mathrm{diag}(z^{-2}, 1, z^{-2}, z^{-2})  \,,
\end{equation}
which gives  a new  matrix $M^{(1)}$, as in  \eqref{M_tilde}, whose leading order term is now diagonalisable. The diagonalisation of the leading term  can be performed using the transformation,
\begin{equation}
	P^{(2)} = \begin{pmatrix}
		0 & -1 & 0 & -1 \\0 & -i{a}/{\Omega} & 0 & i{a}/{\Omega} \\
		1 & 0 & -1 & 0 \\
		1 & 1 & 1 & 1
	\end{pmatrix} \,,
\end{equation}
which yields a matrix $M^{(2)}$ of the form 
\begin{eqnarray}
	M^{(2)} = M_0^{(2)} + M_{1}^{(2)} z^{-1}+  {\cal O}\left( \frac{1}{z^2}\right) \, , \qquad
	M_{0}^{(2)} = \text{diag}(- i \Omega , +i\Omega, -i\Omega, +i\Omega) \, .
\end{eqnarray}
One thus finds four modes propagating at speed $c=1$, two  ingoing  and two outgoing modes.  {We expect them to be associated with the scalar and polar gravitational degrees of freedom.}

\medskip

In order to discriminate between the scalar and gravitational modes, it is useful to pursue the diagonalisation up to the next-to-leading order.
This can be done by following, step by step, the algorithm of \cite{Langlois:2021xzq}, which leads us to introduce the successive matrices $P^{(3)}$ and $P^{(4)}$,
\begin{equation*}
	P^{(3)} = I_4 + \frac{i}{2\Omega z} \begin{pmatrix}
		0 & 0 & -1 & -2 \\
		0 & 0 & 1 & -(1 + 2 \Omega^2) \\
		1 & 2 & 0 & 0 \\
		-1 & 1 - 2\Omega^2 & 0 & 0
	\end{pmatrix} \,, \qquad
	P^{(4)} = \begin{pmatrix}
		-3a + b & 1 & 0 & 0 \\
		1 & 0 & 0 & 0 \\
		0 & 0 & -3a + b^* & 1 \\
		0 & 0 & 1 & 0
	\end{pmatrix} \,,
\end{equation*}
with the complex coefficient $b$ defined by
\begin{equation}
	b = -\frac12 + \frac{\varepsilon^2}{24} i \Omega (1 - 3 \Omega^2 - 36 i \Omega) \,.
\end{equation}
Hence, we obtain a new vector $\X^{(4)}$ whose corresponding matrix $M^{(4)}$ is given by
\begin{eqnarray}
	M^{(4)} & = & \text{diag}(-i\Omega, -i\Omega, i\Omega, i\Omega) \nonumber \\
	&+&  \frac{1}{z} \text{diag}\qty(-1-i\Omega, 3 - i\Omega \qty(1 + {\varepsilon^2}/{3}), -1+i\Omega, 3 + i\Omega \qty(1 + {\varepsilon^2}/{3})) + {\cal O}\left( \frac{1}{z^2}\right)\,,
\end{eqnarray}
up to order  $\varepsilon^2$. As a consequence, we can now easily integrate the equation for  $Y^{(4)}$  up to sub-leading order when $z \gg 1$ (up to $\varepsilon^2$) and we obtain
\begin{eqnarray}
	{}^t Y^{(4)} \simeq \begin{pmatrix}
		c_- \, \mathfrak{s}^{\infty}_-(z) \, , \quad
		d_- \, \mathfrak{g}^{\infty}_-(z) \, , \quad 
		c_+ \, \mathfrak{s}^{\infty}_+(z) \, , \quad
		d_+ \, \mathfrak{g}^{\infty}_+(z) 
	\end{pmatrix} \, ,
\end{eqnarray}
where $c_\pm$ and $d_\pm$ are integration constants while
\begin{align}
	\mathfrak{g}^{\infty}_\pm(z) \simeq e^{\pm i  \Omega z} z^{3 \pm i\Omega \qty(1+ {\varepsilon^2}/{3})} = e^{\pm i z_*} \,, \qquad
	\mathfrak{s}^{\infty}_\pm(z)  \simeq e^{\pm i  \Omega z} z^{-1 \pm i\Omega }  \,.
	\label{eq:asymp-behav-infty-aEGB-polar}
\end{align}
The two modes $\mathfrak{g}^{\infty}_\pm$  follow the same behaviour as the axial modes obtained in \eqref{eq:asymp-mat-axial-4EGB}: those can be dubbed  gravitational modes, while the other two modes $\mathfrak{s}^{\infty}_\pm$  correspond to scalar modes.

We can then determine the behaviour of the metric perturbations $K$, $\chi$, $H_1$ and $H_0$ by combining the matrices $P^{(i)}$, with $i = 1 , \dots, 4$ as
\begin{eqnarray}
	Y = P \,  Y^{(4)} \quad \text{with} \;\; \qquad P =
	P^{(1)} P^{(2)} P^{(3)} P^{(4)} \,  .
\end{eqnarray} 
with the leading order terms of each coefficient of $P$ given by
\begin{eqnarray}
	P \simeq \frac{1}{z^2}\begin{pmatrix}
		-{1}  & -\dfrac{1}{2i z \Omega} & -{1}  & \dfrac{1}{2i z \Omega}  \\
		-\dfrac{i a z^2}{\Omega} & \dfrac{a z}{2 \Omega^2 } & \frac{i a z^2}{\Omega} &   \dfrac{a z}{2 \Omega^2 }  \\
		{-3a+b}  & {1} & {3a-b} & - {1} \\
		{-3a-b} & {1} & -{(3a+b^*)}  & - {1}
	\end{pmatrix} \,.
\end{eqnarray}
Hence, the metric and the scalar perturbations are non-trivial linear combinations of the so-called gravitational and scalar modes. This shows that the metric and the scalar variables are dynamically entangled.  

%

\subsubsection{Near the horizon}
{The asymptotic behaviour of polar perturbations near the horizon is technically more complex to analyse than the previous case because
	we need more steps to \enquote{diagonalise} the matrix M and then to integrate asymptotically the system for the perturbations}. However, the procedure is straightforward  following the algorithm presented in \cite{Langlois:2021xzq}. For this reason, we do not give the details of the calculation but instead  present the final result.

After several changes of variables, one obtains a first order differential system satisfied by a vector $\tilde Y$ whose corresponding matrix $\tilde{M}$ is of the form 
	\begin{eqnarray}
		\tilde{M} = \frac{1}{z-1} \tilde{M}_{{}_{-1}} + {\cal O}\left( 1\right) \, , 
	\end{eqnarray}
where the leading order term $M_{_{-1}}$ is, up to $\varepsilon^2$, given by
\begin{equation}
	\tilde{M}_{{}_{-1}} = \text{diag} \left[ - i \Omega \qty(1 - \frac{21}{10}\varepsilon^2),+i\Omega \qty(1 - \frac{21}{10}\varepsilon^2),-i\Omega \qty(1 - \frac{21}{10}\varepsilon^2),+i\Omega \qty(1 - \frac{21}{10}\varepsilon^2) \right]  + \mathcal{O}(\varepsilon^3)\; . 
\end{equation}
One recognises that the coefficients of $M_{_{-1}}$ correspond to the leading order term in the asymptotic expansion of $\pm i \Omega z_*$ around $z = 1$, given in \eqref{eq:z-tortoise-horizon}. Indeed, we see that
\begin{eqnarray}
	\tilde M =  i \Omega \, \frac{\dd z_*}{\dd z} \, \text{diag} \left( -1,+1,-1,+1\right) + {\cal O}(1)\, ,
\end{eqnarray}
and then integrating the equation for $\tilde Y$  becomes trivial as
\begin{eqnarray}
	\frac{\dd \tilde Y}{\dd z_*} \simeq    \, \text{diag} \left( -i \Omega ,+i \Omega ,-i \Omega ,+i \Omega  \right) \tilde Y \, ,
\end{eqnarray}
which leads to the solution
\begin{eqnarray}
	{}^t \tilde Y = (c_-\mathfrak{p}^{1}_-(z)\, , \quad c_+\mathfrak{p}^{1}_+(z)\, , \quad d_-\mathfrak{p}^{1}_-(z)\, , \quad d_+ \mathfrak{p}^{1}_+(z)\, ) \, ,
\end{eqnarray}
where $c_\pm$ and $d_\pm$ are integration constants, and we introduced the polar modes (up to $\varepsilon^2$),
\begin{align}
	\mathfrak{p}^{1}_\pm(z) \simeq e^{\pm i \Omega z_*} = (z-1)^{\pm i \Omega (1 - 21  \varepsilon^2/10)}  \, .
	\label{eq:asymp-behav-hor-aEGB-polar}
\end{align}
Several remarks are in order. First,
exactly as in the analysis of the asymptotics at infinity, one cannot  discriminate between the gravitational mode and the scalar mode 
at  leading order since they are equivalent at this order.  Going to next-to-leading  orders would be needed in order to further characterise each mode.   Then, computing the behaviour of each mode at the horizon in terms of the metric perturbation functions, in a similar way to what was done at spatial  infinity, is possible but not enlightening since the expressions are very involved. 
Finally, notice that the results above \eqref{eq:asymp-behav-infty-aEGB-polar} and \eqref{eq:asymp-behav-hor-aEGB-polar} are consistent with the behaviours found in \cite{Blazquez-Salcedo:2016enn}, as one can see in their equation (17), and \cite{Minamitsuji:2022mlv} as one can see in their equations (6.62) and (6.63).

\section{4d Einstein-Gauss-Bonnet black hole}
\label{section_4dGB}

In this section, we  study another modified theory of gravity that  involves the Gauss-Bonnet invariant  $\gb$ defined in \eqref{GBinvariant}. Its action is given by
\begin{equation}
	S[g_{\mu\nu}, \phi] = \int \dd[4]{x} \sqrt{-g} \left(R + \alpha(\phi \gb + 4 E^{\mu\nu}\phi_\mu\phi_\nu - 4 X \square\phi + 2X^2)\right) \,,
	\label{eq:action-4GB}
\end{equation}
{where $\alpha$ is a constant and  $E^{\mu\nu}$  the Einstein tensor.  This action can be obtained as the $4D$ limit}, in some specific sense, of the $D$-dimensional Einstein-Gauss-Bonnet action \cite{Lu:2020iav}.
As for Einstein-scalar-Gauss-Bonnet theories, this theory also belongs to  degenerate scalar-tensor theories. It can be recast into a  Horndeski theory with the following functions (see Appendix~\ref{app:equivalence-EGB-Horndeski}):
\begin{equation}
	\label{functions_4dGB}
	P(X) = 2\alpha X^2 \,,\quad Q(X) = -4\alpha X \,,\quad F(X) = 1 - 2\alpha X \qq{and} G(X) = -4\alpha \ln X \,.
\end{equation}
We will also assume that $\alpha > 0$, otherwise $|\alpha|$ is constrained to be extremely small \cite{Charmousis:2021npl}. 

\subsection{Background solution}

Let us now consider static spherically symmetric solutions of  the form
\begin{equation}
	\dd{s}^2 = - A(r) \dd{t}^2 + \frac{1}{A(r)} \dd{r}^2 + r^2\dd{\Omega}^2 \qq{and} \phi = \phi(r) \,.
\end{equation}
By solving the equations of motion for the metric and the scalar field derived from the action \eqref{eq:action-4GB}, one can find a simple analytical solution, as discussed in \cite{Lu:2020iav,Hennigar:2020lsl}. The metric function $A$ is given by
\begin{equation}
	A(r) = 1 + \frac{r^2}{2\alpha} \left(1 - \sqrt{1 + \frac{4\alpha\mu}{r^3}}\right)=1-  \frac{2\mu/r}{1+\sqrt{1+\frac{4\alpha\mu}{r^3}}}\,.
	\label{eq:metric-function-r}
\end{equation}
This reduces to the Schwarzschild metric in the limit   $\alpha \rightarrow 0$,   the parameter $\mu$  corresponding to twice the black hole mass in this limit. 

{If $\mu^2 < 4\alpha$, the solution is a naked singularity and is therefore of no interest. If $\mu^2 \geq 4\alpha$,} the solution for the metric describes a black hole and its horizons can be found by solving the equation $A(r) = 0$ for $r$. This gives two roots, the largest one corresponding to the outermost horizon,
\begin{equation}
	\label{rh_mu}
	r_h = \frac{1}{2} \left({\mu}+ {\sqrt{\mu^2-4\alpha}} \right) \, . 
\end{equation}

%
%
%

The equation for the scalar field gives two different branches:
\begin{equation}
	\phi'(r) = \frac{\sigma + \sqrt{A(r)}}{r \sqrt{A(r)}} \qq{with} \sigma = \pm 1\, .
	\label{eq:sol-phi}
\end{equation}
Integrating this equation in the limit where $r$ is large (i.e. $r \gg r_h$), one obtains
\begin{eqnarray}
	\label{scalaratinf}
	\phi(r) \simeq \frac{\mu}{2r} \quad \text{if} \; \sigma=-1 \, , \qquad
	\phi(r) \simeq 2 \ln \left(\frac{r}{\mu} \right) \quad \text{if} \; \sigma=+1 \, .
\end{eqnarray}
Hence, the branch $\sigma=+1$ leads to a divergent behaviour of the scalar field at spatial infinity. In this branch, moreover,  $\phi$ does not vanish when the black hole mass goes to zero  and we will see later that the perturbations feature also a pathological
behaviour. For these reasons, we will mostly 
restrict our analysis to the  branch $\sigma=-1$.

\medskip
In the following, it will be convenient to use the dimensionless quantities 
\begin{equation}
	z = \frac{r}{r_h} \qq{and} \beta = \frac{\alpha}{r_h^2} \,.
\end{equation}
According to these definitions and \eqref{rh_mu}, one can replace $\mu$ by $(1+\beta) r_h$. Note that
\begin{equation}
	0 \leq \beta = \frac{\mu - r_h}{r_h} \leq 1 \,,
	\label{eq:bounds-alpha-rp}
\end{equation}
as $0 \leq r_h \leq \mu$. One can notice that both bounds can be reached: the case $\beta = 0$ is the GR limit, while the case $\beta=1$ is an extremal black hole, as both horizons merge into one located at $r_h = \sqrt{\alpha}$. The parameter $\beta$ is therefore similar to the extremality parameter $Q/M$ for a charged black hole, and it is interesting to use it instead of $\alpha$ when studying the present family of black hole solutions.

Moreover, the outermost horizon is now at $z = 1$ and the new metric function is
\begin{equation}
	\label{A_z}
	A(z) = 1 + \frac{z^2}{2\beta} \qty(1 - \sqrt{1 + \frac{4\beta(1+\beta)}{z^3}}) = 1 - \frac{2(1+\beta)}{z \left(1+\sqrt{1 + \frac{4\beta(1+\beta)}{z^3}}\right)}\,.
\end{equation}
Since $\phi'$ depends on $\sqrt{A}$, as shown in \eqref{eq:sol-phi}, it is also convenient to introduce the new function
\begin{equation}
	\label{f_z}
	f(z) = \sqrt{A(z)} \,.
\end{equation}

\subsection{Axial modes:  the first order system}
\label{subsec:schrodinger-eq}

The dynamics of axial modes is described by a canonical system of the form \eqref{eq:system-axial-canonical}.  Substituting (\ref{functions_4dGB}),  (\ref{eq:sol-phi}), (\ref{A_z}), (\ref{f_z}) into the definitions (\ref{Fc}) and (\ref{matrix_coeffs_1}-\ref{matrix_coeffs_2}) and rescaling all dimensionful quantities by the appropriate powers of $r_h$ to make them dimensionless (or, equivalently, working in units where $r_h=1$), one gets the following  expressions for $\mathcal{F}$, $\Gamma$, $\Phi$ and $\Delta$ :
\begin{align}
	\mathcal{F} &
	= \frac{f^2}{z^2} \qty[z^2 + 2\beta(\sigma + f)(\sigma + f - 2z f')]\,,
	\label{Fcal}
	\\
	\label{Gamma_z}
	\Gamma &=   \frac{1}{\mathcal{F} f^2z^2 }\left[z^2 - 2\beta (1 -  f^2) - 4z\beta ff'\right] = \frac{z^4 - 2\beta(1+\beta)z}{\mathcal{F} f^2z^2  [z^2 + 2\beta(1-f^2)]} \,,
	\\
	\Phi &= \frac{\mathcal{F} z^2}{z^2 + 2\beta(1 - f^2)}  \,,\qquad 
	\Delta =  - \frac{\mathcal{F}'}{\mathcal{F} 
	}  \,,\label{eq:expr-canonical-df}
\end{align}
where we have used the explicit definition of $f(z)$ and the expression of its derivative
\begin{equation}
	f'= \frac{f^2 - 1}{z f} + \frac{3(1+\beta)}{2f[z^2 - 2\beta (f^2 - 1)]}
\end{equation}
to obtain a  simplified  expression for $\Gamma$. Here, we have kept the parameter $\sigma$ unfixed: as we can see, it appears in the expression of $\mathcal{F}$ which means that it becomes relevant for the perturbations of the black hole solution. 

%

In the sequel, it will be  convenient to express the quantities \eqref{Fcal}-\eqref{eq:expr-canonical-df} in terms of the following three  functions of $z$:
\begin{align}
	\gamma_1 &= f \qty[z^2 + 2\beta(\sigma + f)(\sigma + f - 2 z f') ] \,,\\
	\gamma_2 &= z^4 - 2\beta(1+\beta)z \,,\\
	\gamma_3 &= z^2 + 2\beta(1-f^2) \,.
\end{align}
A short calculation then leads  to
\begin{equation}
	\label{F_gamma_i}
	\mathcal{F} = \frac{f \gamma_1}{z^2} \,,\quad \Gamma = \frac{\gamma_2}{f^3 \gamma_1 \gamma_3} \qq{and} \Phi = \frac{f \gamma_1}{\gamma_3} \,.
\end{equation}
When we study the perturbations and their asymptotics, it is important to look at the zeros and the singularities of the expressions \eqref{F_gamma_i}. For
this reason, we quickly discuss the zeros of the functions $\gamma_i$. We note that, for $z>0$,  the function $\gamma_3$, explicitly given by 
\begin{equation}
	\gamma_3 = z^2 \sqrt{1 + \frac{4\beta(1+\beta)}{z^3}} \,, 
\end{equation}
is strictly positive  and 
the function $\gamma_2$ vanishes 
  at
\begin{equation}
	z_2 = \qty[2\beta(1+\beta)]^{1/3} \, .
\end{equation}
This root is only relevant in our analysis if it lies outside the horizon, i.e. when  $z_2 > 1$,  which is the case  if $\beta  \geq \beta_c$ 
with
\begin{equation}
	\label{beta_c}
	\beta_c\equiv \frac{\sqrt{3} - 1}{2} \simeq 0.366\, .
\end{equation}
Hence, when $\beta < \beta_c$, $\gamma_2$ remains strictly positive outside the horizon.
{ Let us note that at the special value $\beta=\beta_c$, the zeros of $f$ and $\gamma_2$ coincide.}
Finally, the position of the zeros of $\gamma_1$ depends on the sign of $\sigma$. If $\sigma = -1$, then $\sigma + f \leq 0$ and, since $f' \geq 0$,  the product $(\sigma + f)(\sigma + f - 2zf')$ is always positive, and therefore  $\gamma_1>0$  outside the horizon. By contrast, if $\sigma = +1$, one finds numerically that $\gamma_1$ has a zero $z_1 > 1$. This is another reason (in addition to the behaviour of the scalar field at infinity discussed below \eqref{scalaratinf}) to restrict our analysis to the case
$\sigma=-1$.

Let us summarise. When $\beta < \beta_c$ and $\sigma=-1$, the functions $\gamma_i$ do not vanish outside the horizon and then neither
 of the three functions $\cal F$, $\Gamma$ and $\Phi$ vanishes or has a pole for $z >1$. Near the horizon, these functions behave as follows:
\begin{eqnarray}
	z\to 1 : \; {\cal F} \simeq \frac{6 \beta (1+\beta)}{1+2\beta} \, f \, , \quad
	\Gamma \simeq \frac{(1+2\beta)(1-2\beta-\beta^2)}{6\beta(1+\beta)}\frac{1}{f^3} \, , \quad
	\Phi \simeq  \frac{6 \beta (1+\beta)}{1+2\beta} \, f \, ,
\end{eqnarray}
with 
\beq
\label{f_asympt_1}
f(z)= \sqrt{\frac{1-\beta}{1+2\beta}}\, \sqrt{z-1} +{\cal O}((z-1)^{3/2})\,.
\eeq
At infinity, the behaviour is much simpler as the three functions \eqref{F_gamma_i} are constant and tend to $1$.

\medskip

\subsection{Axial modes: Schr\"odinger-like formulation}
As discussed in \ref{subsection_schroedinger}, the axial perturbations obey the Schr\"odinger-like  equation 
\begin{equation}
	\dv[2]{\hat\X_1}{z_*} + \left(\frac{\Omega^2}{c^2(z)}-V(z) \right) \hat\X_1= 0 \,,
\end{equation}
where $\dv*{z}{z_*} = n(z)$ and $c^2 = 1/n^2\Gamma$, while the potential $V(z)$ is given by (\ref{generalpotentialtext}).  The condition $\beta<\beta_c$ together with the choice $\sigma=-1$ ensures that $c^2>0$ everywhere outside the horizon.

\begin{figure}[!htb]
	\begin{subfigure}{0.45\textwidth}
		\centering
		\includegraphics{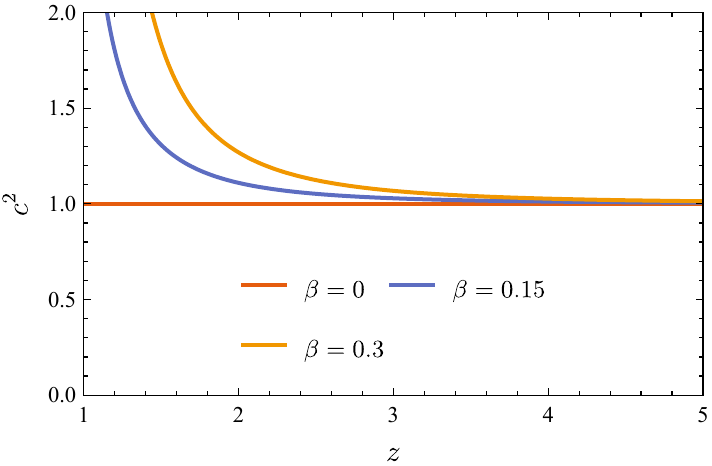}
		\caption{Squared speed}
	\end{subfigure}
	\begin{subfigure}{0.45\textwidth}
		\centering
		\includegraphics{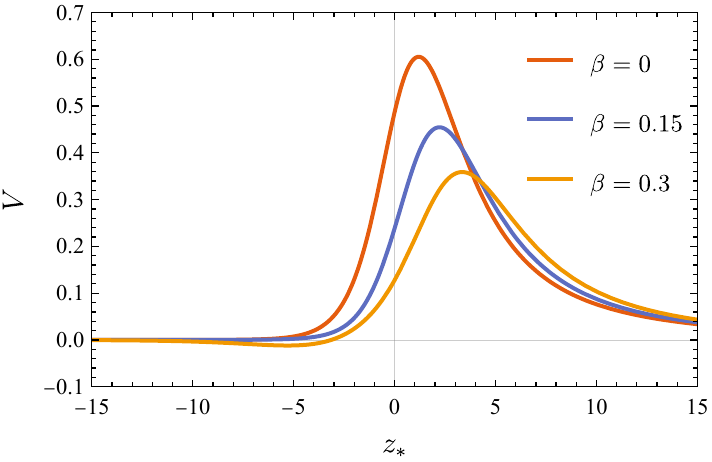}
		\caption{Potential}
	\end{subfigure}
	\caption{Plot of the squared speed $c^2$ and the potential $V$ for $\lambda=2$. We choose the integration constant in the computation of $z_*$ through the same procedure as explained in Fig.~\eqref{fig:c-pot-aEGB}.}
	\label{fig:c-pot-orig}
\end{figure}

A natural choice for $n$ is $n(z) = A(z) = f^2(z)$, in which case $z_*$ is the analog of  the Schwarzschild tortoise coordinate. With this choice, one finds, according to (\ref{F_gamma_i}),
\begin{equation}
	c^2 = \frac{\gamma_1\gamma_3}{f \gamma_2} \,.
	\label{eq:speed-tortoise}
\end{equation}
The potential is then given by
\begin{eqnarray}
	V(z) &= &\frac{z^2 A (\kappa_1 + A \kappa_2)}{\gamma_2^2 \gamma_3^4} \,,\quad\text{with} \label{eq:potential-tortoise} \\
	\kappa_1 &= & 2 (\lambda +1) z^{12}-3 (\beta +1) z^{11}-2 \beta  (\beta +1) (2 \lambda -7) z^9-18 \beta  (\beta +1)^2 z^8  \nonumber\\ 
	&-&24 \beta ^2 (\beta +1)^2 (\lambda +1) z^6 + 54 \beta ^2 (\beta +1)^3 z^5 + 4 \beta ^3 (\beta +1)^3 (20 \lambda -7) z^3 \nonumber \\
	& -& 12 \beta ^3 (\beta +1)^4 z^2-8 \beta ^4 (\beta +1)^4 (8 \lambda -1) \,,\nonumber\\
	\kappa_2 &= & 30 \beta  (\beta +1) z^9 + 126 \beta ^2 (\beta +1)^2 z^6 + 108 \beta ^3 (\beta +1)^3 z^3 + 12 \beta ^4 (\beta +1)^4 \,. \nonumber
\end{eqnarray}
The propagation speed and the potential for $\lambda=2$ are represented in Fig.~\eqref{fig:c-pot-orig} for three different values of $\beta$, satisfying the condition $\beta<\beta_c$. 
We observe that the propagation speed diverges at the horizon $z=1$, while the potential vanishes at this point. The potential can be negative in some region for  sufficiently large values of $\beta$. It is difficult to study analytically the sign of the potential but one can compute its derivative when $z_*\to -\infty$ and one finds that it remains positive up to some value $\beta_*(\lambda)$. We find that $\beta_*(\lambda=2)\simeq 0.162917$ numerically and that $\beta_*(\lambda\to\infty)=\beta_c$.


\medskip
In terms of the new coordinate $z_*$, the Schr\"odinger-like equation is of the form
\beq
\label{sturm-liouville}
-\dv[2]{\hat\X_1}{z_*} +V(z)  \chi = w(z) \, \Omega^2\hat\X_1  \,,\qquad w=f^4\Gamma\,.
\eeq
The left-hand side of  this equation can be seen as 
an operator  acting on the space of functions that are square-integrable with respect to the measure $w \, \dd{z_*}$. 
It is instructive to  study the asymptotic behaviour of the  solutions of \eqref{sturm-liouville},  near the horizon and at spatial infinity.

Near the horizon, using $\dd{z_*}=\dd{z}/f^2$ and \eqref{f_asympt_1}, one finds
\beq
\label{zstar4D}
z_*\simeq  \frac{1+2\beta}{1-\beta}\ln({z-1}) \; \Longleftrightarrow \; z-1 \simeq e^{\eta z_*} \, , \quad \eta\equiv \frac{1-\beta}{1+2\beta}\ \,,\quad (z\to 1\quad  \text{or} \quad z_*\to -\infty)
\eeq
and  the asymptotic behaviours for the potential and for $w$ are 
\beq
V(z)\simeq C_1 (z-1) , \qquad w(z)\simeq \frac{1-2\beta-2\beta^2}{2\beta\sqrt{(1-\beta)(1+2\beta)}}\sqrt{z-1} \, ,
\eeq
where $C_1$ is a constant. It is immediate to rewrite these asymptotic expressions in terms of $z_*$, using \eqref{zstar4D}.

Near the horizon, for $z_*\to -\infty$, the potential decays faster than the right-hand side of \eqref{sturm-liouville} so that the differential equation takes the form
\beq
-\dv[2]{\hat\X_1}{z_*} +C_1 e^{\eta z_*/2} \hat\X_1 \simeq 0\,,
\eeq
whose solutions are
\beq
\label{solbessel}
\hat\X_1 \simeq A_1 I_0\left(\frac2\eta C_1^{1/2} e^{\eta z_*/4}\right)+A_2 K_0\left(\frac2\eta C_1^{1/2} e^{\eta z_*/4}\right)\,,\qquad (z_*\to -\infty)
\eeq
where $I_0$ and $K_0$ are the modified Bessel functions of order $0$ while $A_1$ and $A_2$ are integration constants. 

Since $I_0(u)\simeq 1$ and $K_0(u)\simeq -\ln u$ when $u\to 0$, the general solution behaves as an affine function of $z_*$ when $z_*\to -\infty$ and is therefore square integrable with respect to the measure $w\, dz_*\simeq e^{\eta z_*/2} dz_*$.  This means that the endpoint $z_* \to -\infty$ is of limit circle type (according to the standard terminology, see e.g. \cite{krallSingularSelfadjointSturmLiouville1988}). {Interestingly, the analysis of the axial modes near the horizon in our case is very similar to that near a naked singularity as discussed in  \cite{Sadhu:2012ur}}. {In contrast with the GR case, none of the two axial modes is  ingoing or outgoing, which means that the 
stability 	analysis  of these perturbations differs from the GR one.}

\medskip

For the other endpoint (at spatial infinity), $z_*\simeq z \to +\infty$,  the asymptotic behaviours of the potential $V$ and the functions $w$, according to \eqref{eq:potential-tortoise} and \eqref{eq:speed-tortoise}, are given by
\beq
V(z)\simeq \frac{2(\lambda+1)}{z^{2}}\,, \qquad w(z)\simeq 1\,,
\eeq
which coincides with the GR behaviour at spatial infinity.
In particular, $V$ goes to zero and $w$ goes to one, so that one recovers the usual combination of ingoing and outgoing modes 
\beq
\hat\X_1\simeq B_1 e^{i\Omega z_*}+B_2  e^{-i\Omega z_*} \,, \qquad (z_*\to +\infty)\,,
\eeq
where $B_1$ and $B_2$ are constant.
If $\Omega$ contains a nonzero imaginary part, then one of the modes is normalisable and then this endpoint is now of limit-point type. 

\medskip

As we have already said previously, the analysis of axial perturbations in this theory is very different from the analysis in GR. The main
reason is that we  no longer have a distinction between ingoing and outgoing modes at the horizon. The choice of the right behaviour
to consider might be guided by regularity properties of the mode. Indeed, if we require the perturbation\footnote{The regularity concerns the metric components themselves and not directly the function $\hat \X_1$. The asymptotic behaviour of the metric components will be given in \eqref{asympaxial4dhor}.} $\hat\X_1$ to be regular 
when $z_*\to -\infty$, then we have to impose $A_2=0$. The problem turns into a Sturm-Liouville problem, which implies that 
$\Omega^2$ is real. A very similar problem has been studied in another context in \cite{Sadhu:2012ur} where the authors showed that
$\Omega^2>0$ when $V>0$, which implies that the perturbations are stable.  Here we can make the same analysis as in \cite{Sadhu:2012ur}, and we expect the stability result to be true at least in the case where $V>0$, i.e. when $\beta$ is sufficiently small, as explained in the discussion below \eqref{eq:potential-tortoise}.

\begin{figure}[!htb]
	\centering
	\includegraphics{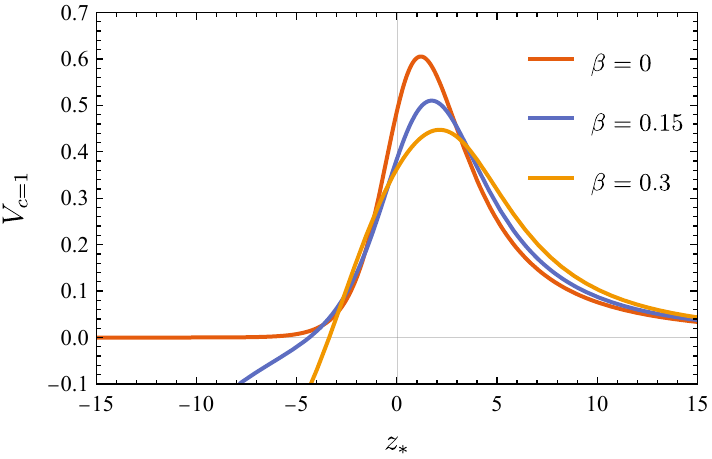}
	\caption{\small{Plot of the potential $V_{c=1}$ for $\lambda=2$.}}
	\label{fig:pot-c1}
\end{figure}

\medskip
Let us close this subsection with a final remark.  It is always possible to use, instead of the tortoise coordinate, a different coordinate $z_*$, for example by choosing $n(z)$ such that $c = 1$ everywhere. This corresponds to the choice
\begin{equation}
	n(z) = \frac{1}{\sqrt{\Gamma}} \,.
\end{equation}
%
%
%
%
In this new frame, the potential is changed and can be written in the form
\begin{equation}
	V_{c=1} = \frac{Q(f)}{16 z^2 f \gamma_1 \gamma_2^3 \gamma_3^5} \,,
\end{equation}
where $Q$ is a polynomial of order 28 of nonzero constant term whose coefficients depend on $z$. This potential is represented on Fig.\eqref{fig:pot-c1} for different values of $\beta$. 


\subsection{Axial modes: first-order asymptotic approach}
\label{subsec:boundary-modes-axial}

In this section, we compute the asymptotic behaviours of $h_0$ and $\hc$ using the first-order system\footnote{The change of variables leading from $\omega$ to $\Omega$ requires to rescale $\hc$ by a factor $r_h$.} for axial perturbations given in \eqref{eq:system-axial-canonical} following the algorithm we developped in \cite{Langlois:2021xzq}. 
Using \eqref{eq:system-axial-canonical}, with \eqref{Fcal}-\eqref{eq:expr-canonical-df}, 
we start by writing this first order system in  the $z$ variable:
\begin{equation}
	\dv{\X}{z} = M(z)\,  \X\,, \qq{with} \X = \begin{pmatrix} h_0 \\ \hc \end{pmatrix} \,.
	\label{eq:first-order-axial}
\end{equation}

At spatial infinity, the matrix $M$ can be expanded as
\begin{equation}
	M(z) = \begin{pmatrix}
		0 & -i\Omega^2 \\
		-i & 0
	\end{pmatrix} + \mathcal{O}\left(\frac{1}{z}\right) \,.
\end{equation}
Therefore, the two components of $\X$  at infinity are immediately found to be a linear combination of the following two modes:
\begin{equation}
	\mathfrak{a}^{\rm \infty}_\pm(z) \simeq e^{\pm i \Omega z} = e^{\pm i \omega r} \,.
\end{equation}
Hence, the asymptotic behaviour of the original metric variables $h_0$ and $\hc$ are given by
\begin{align}
	h_0(z) & \simeq  z \Omega \left[- c_+ e^{i \Omega z} z^{i\Omega(1+\beta)} +  c_- e^{-i \Omega z} z^{ - i\Omega(1+\beta)} \right]\,,\nonumber\\
	\hc(z) & \simeq z \left[ c_+ e^{i \Omega z} z^{ i\Omega(1+\beta)} + c_- e^{-i \Omega z} z^{ - i\Omega(1+\beta)} \right]\,,
\end{align}
where $c_\pm$ are constants. 

Near the horizon, we change variables by setting $x = 1/\sqrt{z-1}$, and study the behaviour, when $x$ goes to infinity, of the system  \eqref{eq:first-order-axial}, rewritten as
\begin{equation}
	\label{Matx}
	\dv{\X}{x} = M_x(x)\,  \X\,, \qq{with}  M_x(x)= - \frac{2}{x^3} M(1 + 1/x^2) \,.
\end{equation}
The  algorithm then enables us to simplify the original system, here up to order $x^{-1}$, using  the transfer matrix $P$ such that
\begin{equation}
	P = \frac{1}{x p_3}\begin{pmatrix}
		p_1 + x p_2 & p_2 \\
		0 & x^2 p_3
	\end{pmatrix} \,,
\end{equation}
with the functions $p_i$ defined by
\begin{align}	
	&p_1 = (1-\beta)^2 \qty(1+2\beta + 6\beta^2) \,,\quad p_2 = 2(1-\beta)^2\beta\sqrt{1+ \beta - 2\beta^2} \,,\nonumber\\
	&p_3 = 2 i (1+ 2 \beta)^2 (1-2\beta(1+\beta)) \,.
\end{align}
The new system is then
\begin{equation}
	\dv{\tilde{\X}}{x} = \tilde{M}_x \, \tilde{\X} \,, \qquad
	\text{with} \quad 
	\tilde{M}_x= \begin{pmatrix}
		0 & 0 \\
		1/x & 0
	\end{pmatrix} + \mathcal{O}\qty(\frac{1}{x^2}) \,.
\end{equation}
Therefore, the solution near the horizon (written in terms of  the original variable $z$) is a linear combination of  two ``modes'', 
\begin{equation}
	\mathfrak{a}^{\rm h}_1(z)  \simeq 1 \qq{and} \mathfrak{a}^{\rm h}_2(z) \simeq -\frac12\ln(z-1) \,.
\end{equation}
Going back to the original variables $h_0$ and $\hc$, we get
\begin{align}
	h_0(z) & \simeq \frac{p_2}{p_3} c_1 + \sqrt{z-1} \qty(\frac{p_1}{p_3} c_1 + \frac{p_2}{p_3} c_2 - \frac{p_2}{2p_3} c_1 \ln(z-1))\,,\nonumber\\
	\hc(z) & \simeq  \frac{1}{\sqrt{z-1}} \qty(c_2 - \frac12 c_1 \ln(z-1))\,,
	\label{asympaxial4dhor}
\end{align}
with $c_1$ and $c_2$ two constants\footnote{We do not call them $c_+$ and $c_-$ as usual here since it is not possible to identify ingoing and outgoing modes.}. This result is consistent with the asymptotic solution we found from the Schr\"odinger-like equation \eqref{solbessel} when we expand the Bessel functions in power series.

\subsection{Polar modes}

In order to compute the asymptotical behaviour of the polar modes, we proceed similarly to section \ref{subsec:boundary-modes-axial}. We start by writing the system as $\dv*{\X}{z} = M(z) \X$, 
with $\X = {}^t\!\begin{pmatrix}K, & \delta\varphi, & H_1, & H_0\end{pmatrix}$ and  $M$ is now a 4-dimensional matrix. We then compute the series expansion of the matrix $M$ at the horizon and at infinity, and apply the algorithm in order to diagonalise the system up to order 
$x^{-1}$
in each case using a change of vector $\X = \tilde{P} \tilde{\X}$. The matrix $\tilde{P}$ being much more involved than in the axial case, we do not give it explicitly here.

\subsubsection{Spatial infinity}

At spatial infinity, the diagonalized matrix $\tilde{M}$ is found to be
\begin{eqnarray}
	\tilde{M}(z) &= &\text{diag}(0,0,-i\Omega,i\Omega) \nonumber \\
	&+&  \frac{1}{z} \text{diag}\left[-5-i\sqrt{\lambda} ,-5+i\sqrt{\lambda}, 1 - i\Omega(1+\beta), 1 + i\Omega(1+\beta)\right]  + \mathcal{O}\left(\frac{1}{z^2}\right) \, .
\end{eqnarray}	
This leads to an asymptotic solution where  $\X$ is a combination of  4 modes, where we recognise 
two polar gravitational modes,
\begin{equation}
	\mathfrak{g}^{\rm \infty}_\pm(z) \simeq e^{\pm i \Omega z} z^{1 \pm i\Omega(1+\beta)}  \, ,
\end{equation}
and identify the other two as scalar modes,
\begin{equation}
	\mathfrak{s}^{\rm \infty}_\pm(z) \simeq  z^{-5\pm i\sqrt{\lambda}} \,.
	\label{eq:asymb-scal-EGB}
\end{equation}
We can recover the behaviour of the metric perturbations $K$, $\delta\varphi$, $H_1$ and $H_0$ which are the components of $\X$ by using the explicit expression of the matrix $\tilde{P}$. After a direct calculation, we find the following behaviour for $\X$ when $z$ goes to infinity:
\begingroup
\renewcommand*{\arraystretch}{2.0}
\begin{eqnarray}
	Y \simeq 
	\begin{pmatrix}
		\dfrac{-i}{\Omega z} & \dfrac{i}{\Omega z} & \dfrac{2 - i\sqrt{\lambda} - \lambda}{\Omega^2} z  &  \dfrac{2 + i\sqrt{\lambda} - \lambda}{\Omega^2} z \\
		\xi & \xi^* & \dfrac{z^6}{24 \beta(1+\beta)} & \dfrac{z^6}{24 \beta(1+\beta)} \\
		-1 & 1 &  \dfrac{2-i\sqrt{\lambda}}{ i \Omega} z^2 &  \dfrac{2+ i\sqrt{\lambda}}{ i \Omega} z^2 \\
		1 & 1 & \dfrac{2}{3}z^3 & \dfrac{2}{3}z^3 
	\end{pmatrix}
	\begin{pmatrix}
		c_+ \mathfrak{g}^{\rm \infty}_+  \\
		c_- \mathfrak{g}^{\rm \infty}_-  \\
		d_+ \mathfrak{s}^{\rm \infty}_+  \\
		d_- \mathfrak{s}^{\rm \infty}_-  
	\end{pmatrix} \, ,
	\label{polarmodes4dGB}
\end{eqnarray}
\endgroup
where 
\begin{equation}
	\xi = \frac{i (53 \lambda -4)}{48 \beta  (\beta +1) \Omega ^3}+\frac{57 \lambda +466}{576 \beta  \Omega^2}+\frac{26327 i (\beta +1)}{2304 \beta  \Omega } -\frac{175 \beta ^2-1954 \beta +175}{36864 \beta } \,.
\end{equation}
and  $c_\pm$ and $d_\pm$ are constants.

This result calls for a few comments.
First, we can see from \eqref{eq:asymb-scal-EGB} that the scalar modes are not propagating at infinity: even though it is possible to identify two branches corresponding to two sign choices, the corresponding modes do not contain exponentials, and the leading order depends on $\lambda$. This implies that there is no choice of $z_*$ such that $\mathfrak{s}^\infty_\pm (z_*) \simeq e^{\pm i z_* / c_0}$, with $c_0$ a constant speed independant of $\lambda$. Such a behaviour for scalar modes leads to the conclusion that defining quasinormal modes of the scalar sector in the usual way (through outgoing boundary conditions at infinity) for this solution is not possible.

Second, one can compare the asymptotic behaviour of the scalar modes with what is obtained by considering only scalar perturbations onto a fixed background; this is done in Appendix \ref{app:asymp-behav-scalar} and we see that the two 
behaviours
are very similar, even though they slightly differ. 
Third, one can observe that the 4-dimensional matrix above \eqref{polarmodes4dGB} is ill-defined in the GR limit where $\beta \to 0$. In fact, the second line of the matrix tends to infinity in this limit. This could be expected, since in that limit there is no degree of freedom associated 
with
 the scalar perturbation, which is obtained precisely from  the second line of the matrix. One could solve this problem by setting $\chi = \beta \, \delta\varphi$ and considering the vector ${}^t\!\begin{pmatrix}K, & \chi, & H_1, & H_0\end{pmatrix}$, similarly to what was done for the EsGB solution in \eqref{eq:def-chi-EsGB}.
 
\subsubsection{Near the horizon}
Near the horizon, we use the variable $x = 1/\sqrt{z-1}$, as for axial modes. Using the algorithm, we find a change of 
vector  
$Y= \tilde P \tilde Y$ such that the associated matrix, that we denote $\tilde{M}_x$ exactly as in \eqref{Matx},  is diagonal and  is explicitly given by
\begin{equation}
	\tilde{M}_x = \frac{1}{x} \, \text{diag}(-1,0,0,2) + \mathcal{O}\qty(\frac{1}{x^2}) \,.
\end{equation}
Solving the first order system is then immediate and the asymptotic expressions of the components of the 4-dimensional vector $\tilde Y$ (written as functions of $z$) are combinations  of  the four modes 
\begin{equation}
	\mathfrak{g}^{\rm h}_1(z) \simeq 1 \,,\quad \mathfrak{g}^{\rm h}_2(z)  \simeq \frac{1}{z-1} \,,\quad \mathfrak{s}^{\rm h}_1(z)  \simeq 1 \qq{and} \mathfrak{s}^{\rm h}_2(z)  \simeq \sqrt{z-1} \,.
\end{equation}
We have named two of these modes $\mathfrak{s}_i$ 
(for \enquote{scalar})
because they contain  a nonzero $\delta \varphi$ contribution, as can be seen by expressing these modes in terms of the original perturbative quantities, using the explicit expression for the matrix $\tilde{P}$ provided by the algorithm\footnote{One can also see from \eqref{Ypolar4dGB} that $\delta\varphi$ is a combination of only these two modes at the horizon, which strengthens this denomination.}. Indeed, the relation between each  of the above 
modes
and  the initial perturbations is given by 
\begin{eqnarray}
	Y \simeq
	\begin{pmatrix}
		\dfrac{\zeta_1}{\sqrt{z-1}} &  \zeta_2 \sqrt{z-1} & \zeta_4 \sqrt{z-1}  & \zeta_6 \sqrt{z-1} \\
		0 & 0 & 1 & \sqrt{z-1} \\
		\dfrac{1}{z-1} & 1& 0 & \zeta_7 \sqrt{z-1} \\
		0 & \dfrac{\zeta_3}{\sqrt{z-1}}& \dfrac{\zeta_5}{\sqrt{z-1}} & \zeta_8 
	\end{pmatrix}
	\begin{pmatrix}
		c_1 \, \mathfrak{g}^{\rm h}_1  \\
		c_2 \, \mathfrak{g}^{\rm h}_2 \\
		d_1 \, \mathfrak{s}^{\rm h}_1  \\
		d_2 \, \mathfrak{s}^{\rm h}_2  
	\end{pmatrix} \, ,
	\label{Ypolar4dGB}
\end{eqnarray}
where $c_i$ and $d_i$ are integration constants while  $\zeta_i$ are constants whose expressions are {given explicitly in Appendix \ref{app:expr-xis}}.

This behaviour is similar to what we have obtained for the axial perturbations. One cannot exhibit ingoing and outgoing modes: instead, the perturbations have non-oscillating behaviours at the horizon. 

\section{Conclusion}

In this work, we have studied the linear perturbations about black hole solutions in the context of two families of gravity theories involving a Gauss-Bonnet term in the action. In order to do so we have extended our previous work to the case of Horndeski theories with a cubic dependence on second derivatives of the scalar field, since the Gauss-Bonnet models studied here can be recast in the form of scalar-tensor theories (we show explicitly, in Appendix \ref{app:equivalence-EGB-Horndeski}, the equivalence between the  Lagrangians with the Gauss-Bonnet term and the corresponding scalar-tensor Lagrangians following what has been done in \cite{colleauxRegularBlackHole2019}).

For a general shift-symmetric Horndeski theory,  we have written the expression for the equations of motion of the axial perturbations about any static spherically symmetric background in a simple and compact form. The axial perturbations represent a single degree of freedom and their dynamics can be described either by a two-dimensional first-order (in 
radial
derivatives) system or by a Schr\"odinger-like second-order equation, 
associated with
a potential and a propagation speed. By contrast, the polar modes,
which
 describe the coupled  even-parity gravitational degree of freedom and the scalar 
 field
 degree of freedom, 
 are
 characterized by a four-dimensional first-order system. We then apply this general formalism to the two models considered here.

For Einstein-scalar-Gauss-Bonnet theories, one difficulty is that there is no exact background black hole solution. The solution can be computed numerically  or analytically in a perturbative expansion. We have followed the second approach here, following \cite{Julie:2019sab} and providing some details about the calculation of the lowest order metric terms. We have then studied the perturbations, up to second order in the small expansion parameter (related to the coupling of the Gauss-Bonnet term). We have tackled the axial modes using both the Schr\"odinger reformulation and the first-order system approach, thus cross-checking our results. As for polar modes, we have applied our algorithm to determine their asymptotic behaviours. We have found that both axial and polar modes have a rather standard behaviour. In particular, one can immediately see the existence of ingoing and outgoing modes at both boundaries and one can easily distinguish in most 
cases gravitational and scalar degrees of freedom at the boundaries.

In the last part of this work, we have  considered the perturbations of the 4dGB black hole solution of \cite{Lu:2020iav}, for the first time to our knowledge. The Schr\"odinger reformulation of the equations of motion for the axial modes is characterised by the unusual property that the propagation speed diverges at the horizon (using the tortoise coordinate as radial coordinate), even if the potential vanishes in this limit. We also find a critical value $\beta_c$ for the coupling beyond which the square of the propagation speed becomes negative. Studying the case $\beta<\beta_c$, we have found that the asymptotic behaviour at spatial infinity is very similar to that of Schwarzschild but the modes are very peculiar near the horizon. These results are confirmed by our first-order approach.

Moreover, concerning both polar and axial perturbations, it is not possible to apply the usual classification of modes into ingoing/outgoing categories near the horizon. Furthermore, we prove that the scalar modes have a leading order behaviour at infinity that strongly depends on the angular momentum, which seems to imply that no scalar waves propagate at infinity.

In summary, we have illustrated in this work how our formalism can be used in a straightforward and systematic way to study the asymptotic behaviours of the perturbations about a black hole solution in a large family of scalar-tensor theories. When the perturbations are well-behaved, this is a useful starting point for the numerical computation of the quasi-normal modes. By contrast, if the perturbations are ill-behaved, it indicates that the solution or even the underlying gravitational theory might be pathological. In this sense, our general formalism can be used as an efficient diagnostic of the healthiness of some modified gravity theory, or at least the viability of some associated black hole solutions. 

\acknowledgements{We would like to thank Eugeny Babichev, Christos Charmousis, F\'elix Juli\'e and Antoine Leh\'ebel for very instructive discussions, and especially Christos Charmousis for suggesting to consider the 4dGB black hole. We also thank Leo Stein for technical advice concerning Mathematica. }


\appendix

\section{Scalar Einstein-Gauss-Bonnet theory as cubic Horndeski}
\label{app:equivalence-EGB-Horndeski}

In this Appendix, we show that a coupling between a scalar field and the Gauss-Bonnet term gives, in 4-dimensional spacetimes, a cubic Horndeski theory. We use the expression of the Gauss-Bonnet term as a total derivative given in \cite{colleauxRegularBlackHole2019} and we reproduce the proof of this reference in a simpler case here. Our result was already proven in \cite{Kobayashi:2011nu}, but was obtained in that case only from the equations of motion. The computation we present here is made at the level of the action.

Let us study the action
\begin{equation}
	S_\text{GB}[g_{\mu\nu}, \phi] = \int \dd[4]{x} \sqrt{-g} f(\phi) \mathcal{G} \,,
	\label{eq:action-GB}
\end{equation}
with $\mathcal{G}$ the Gauss-Bonnet invariant defined by
\begin{equation}
	\mathcal{G} = R_{\mu\nu\rho\sigma} R^{\mu\nu\rho\sigma} - 4 R_{\mu\nu} R_{\mu\nu} + R^2 \,.
\end{equation}
In 4 dimensions, the Lagrangian density $\sqrt{-g}\,  \mathcal{G}$ is a total derivative: therefore integration by parts should allow us to recover a scalar-tensor action from \eqref{eq:action-GB}. It is proven in \cite{Crisostomi:2017ugk} that the Einstein-Gauss-Bonnet action completed with a kinetic term for the scalar field contains only one scalar degree of freedom. It is therefore expected that the action \eqref{eq:action-GB} can be written as a specific case of \eqref{eq:generic-quintinc-Horndeski}.

We use the expression of $\mathcal{G}$ as a total derivative given in \cite{colleauxRegularBlackHole2019}: introducing an arbitrary field $\phi$, one has
\begin{equation}
	\mathcal{G} = -2 \delta^{\mu\nu\alpha\beta}_{\sigma\rho\lambda\delta} \nabla^\delta \qty[\frac{\tensor{\phi}{_\alpha^\lambda} \phi_\beta}{X} \qty(\tensor{R}{_\mu_\nu^\sigma^\rho} + \frac43 \frac{\tensor{\phi}{_\mu^\sigma} \tensor{\phi}{_\nu^\rho}}{X})] \,,
	\label{eq:G-total-derivative-Colleaux}
\end{equation}
where we introduced the tensor
\begin{equation}
	\delta^{\mu\nu\alpha\beta}_{\sigma\rho\lambda\delta} = - \varepsilon^{\mu\nu\alpha\beta} \varepsilon_{\sigma\rho\lambda\delta}  \,.
\end{equation}
The idea of the proof done in \cite{colleauxRegularBlackHole2019} is to generate Riemann terms by using the commutation of covariant derivatives acting on $\phi^\mu$, using the formula
\begin{equation}
	\comm{\nabla_\mu}{\nabla_\nu} \phi^\rho = \tensor{R}{^\rho_\lambda_\mu_\nu} \phi^\lambda \,.
	\label{eq:commutation-nabla}
\end{equation}
In order to obtain the squared Riemann terms present in $\mathcal{G}$, one searches an expression of $\mathcal{G}$ in the schematic form $\nabla_\mu  (\phi_{\nu\rho} \phi_\sigma R_{\lambda\delta\alpha\beta})$. The action of the covariant derivative on $\phi_{\nu\rho}$ will lead to a squared Riemann term as expected. Recovering the Gauss-Bonnet will require antisymmetrization, since one has
\begin{equation}
	\mathcal{G} = \frac{1}{4} \delta^{\mu\nu\alpha\beta}_{\sigma\rho\lambda\delta} \tensor{R}{_\mu_\nu^\sigma^\rho} \tensor{R}{_\alpha_\beta^\lambda^\delta} \,.
	\label{eq:def-GB-Lovelock}
\end{equation}
We will therefore contract the expression with the fully antisymmetric tensor. However, several new terms will be created when the covariant derivative acts on the other parts of the expression: specific tuning of prefactors in front of these terms will be required to make sure only the Gauss-Bonnet invariant is left in the end.

We reproduce the proof in \cite{colleauxRegularBlackHole2019} in the specific case of 4-dimensional spacetime. We start from the generic Lagrangian
\begin{equation}
	L = \delta^{\mu\nu\alpha\beta}_{\sigma\rho\lambda\delta} \nabla^\delta \qty[a_0 \frac{\tensor{\phi}{_\alpha^\lambda} \phi_\beta}{X} \tensor{R}{_\mu_\nu^\sigma^\rho} + a_1 \frac{\tensor{\phi}{_\alpha^\lambda} \phi_\beta}{X^2} \tensor{\phi}{_\mu^\sigma} \tensor{\phi}{_\nu^\rho} ] \,,
\end{equation}
where $a_0$ and $a_1$ are constants. By expanding the covariant derivative $\nabla^\delta$ in $L$, one obtains
\begin{multline}
	L = a_0 \delta^{\mu\nu\alpha\beta}_{\sigma\rho\lambda\delta} \qty[\frac{\nabla^\delta\tensor{\phi}{_\alpha^\lambda}}{X} \phi_\beta \tensor{R}{_\mu_\nu^\sigma^\rho} + \frac{\tensor{\phi}{_\alpha^\lambda}\tensor{\phi}{_\beta^\delta}}{X} \tensor{R}{_\mu_\nu^\sigma^\rho} - \frac{2}{X^2} \tensor{\phi}{_\alpha^\lambda}\tensor{\phi}{_\kappa^\delta} \phi^\kappa \phi_\beta \tensor{R}{_\mu_\nu^\sigma^\rho}] 
	\\
	+ a_1 \delta^{\mu\nu\alpha\beta}_{\sigma\rho\lambda\delta} \qty[ \frac{3}{X^2} \qty(\nabla^\delta\tensor{\phi}{_\alpha^\lambda}) \phi_\beta \tensor{\phi}{_\mu^\sigma}\tensor{\phi}{_\nu^\rho} + \frac{\tensor{\phi}{_\alpha^\lambda}\tensor{\phi}{_\beta^\delta}\tensor{\phi}{_\mu^\sigma}\tensor{\phi}{_\nu^\rho}}{X^2} - \frac{4}{X^3} \phi^\kappa \phi_\beta \tensor{\phi}{_\kappa^\delta}\tensor{\phi}{_\alpha^\lambda}\tensor{\phi}{_\mu^\sigma}\tensor{\phi}{_\nu^\rho} ] \,,
\end{multline}
by regrouping terms that are equal under contraction with the totally antisymmetric tensor. One notices that the covariant derivatives of the Riemann tensors disappear by application of the second Bianchi identities. By using the first Bianchi identities and \cref{eq:commutation-nabla}, one obtains
\begin{equation}
	2 \delta^{\mu\nu\alpha\beta}_{\sigma\rho\lambda\delta} \nabla^\delta\tensor{\phi}{_\alpha^\lambda} = - \delta^{\mu\nu\alpha\beta}_{\sigma\rho\lambda\delta} \tensor{R}{^\lambda^\delta_\alpha_\kappa} \phi^\kappa \,,
\end{equation} 
which allows us to write 
\begin{equation}
	L = a_0 \qty[-\frac{1}{2X} \Omega_{2,0} - \frac{2}{X^2} \Omega_{3,1} + \frac{1}{X} \Omega_{1,1}] + a_1 \qty[-\frac{3}{2X^2} \Omega_{2,1} + \frac{1}{X^2} \Omega_{1,2} - \frac{4}{X^3} \Omega_{3,2}] \,, 
	\label{eq:expr-L-Omega}
\end{equation}
where the functions $\Omega_{i,j}$ are defined in \cite{colleauxRegularBlackHole2019} as
\begin{align}
	&\Omega_{1,0} = \delta^{\mu\nu\alpha\beta}_{\sigma\rho\lambda\delta} \tensor{R}{_\mu_\nu^\sigma^\rho} \tensor{R}{_\alpha_\beta^\lambda^\delta} \,,\quad &&\Omega_{1,2} = \delta^{\mu\nu\alpha\beta}_{\sigma\rho\lambda\delta} \tensor{\phi}{_\mu^\sigma} \tensor{\phi}{_\nu^\rho} \tensor{\phi}{_\alpha^\lambda} \tensor{\phi}{_\beta^\delta} \,,\nonumber\\
	&\Omega_{1,1} = \delta^{\mu\nu\alpha\beta}_{\sigma\rho\lambda\delta} \tensor{R}{_\mu_\nu^\sigma^\rho} \tensor{\phi}{_\alpha^\lambda} \tensor{\phi}{_\beta^\delta} \,,\quad &&\Omega_{3,1} = \delta^{\mu\nu\alpha\beta}_{\sigma\rho\lambda\delta} \phi_\kappa \phi^\lambda \tensor{\phi}{_\alpha^\kappa} \tensor{R}{_\mu_\nu^\sigma^\rho}  \tensor{\phi}{_\beta^\delta} \,,\nonumber\\
	&\Omega_{2,0} = \delta^{\mu\nu\alpha\beta}_{\sigma\rho\lambda\delta} \phi_\kappa \phi^\rho \tensor{R}{_\mu_\nu^\sigma^\kappa} \tensor{R}{_\alpha_\beta^\lambda^\delta} \,,\quad &&\Omega_{3,2} = \delta^{\mu\nu\alpha\beta}_{\sigma\rho\lambda\delta} \phi_\kappa \phi^\sigma \tensor{\phi}{_\mu^\kappa} \tensor{\phi}{_\nu^\rho} \tensor{\phi}{_\alpha^\lambda} \tensor{\phi}{_\beta^\delta} \,,\nonumber\\
	&\Omega_{2,1} = \delta^{\mu\nu\alpha\beta}_{\sigma\rho\lambda\delta} \phi_\kappa \phi^\rho \tensor{R}{_\mu_\nu^\sigma^\kappa} \tensor{\phi}{_\alpha^\lambda} \tensor{\phi}{_\beta^\delta}  \,.
	\label{eq:def-Omegas}
\end{align}
One can then prove the following identities relating the functions $\Omega_{i,j}$:
\begin{align}
	&X \Omega_{1,0} - 4 \Omega_{2,0} = \delta^{\mu\nu\alpha\beta\gamma}_{\sigma\rho\lambda\delta\kappa} \phi_\mu \phi^\sigma \tensor{R}{_\nu_\alpha^\rho^\lambda} \tensor{R}{^\delta^\kappa_\beta_\gamma} = 0 \,,\nonumber\\
	&X \Omega_{1,1} - 2 \Omega_{2,1} - 2 \Omega_{3,1} = \delta^{\mu\nu\alpha\beta\gamma}_{\sigma\rho\lambda\delta\kappa} \phi_\mu \phi^\sigma \tensor{\phi}{_\nu^\rho} \tensor{\phi}{_\alpha^\lambda}\tensor{R}{^\delta^\kappa_\beta_\gamma} = 0 \,,\nonumber\\
	&X \Omega_{1,2} - 4 \Omega_{3,2} = \delta^{\mu\nu\alpha\beta\gamma}_{\sigma\rho\lambda\delta\kappa} \phi_\mu \phi^\sigma \tensor{\phi}{_\nu^\rho} \tensor{\phi}{_\alpha^\lambda} \tensor{\phi}{_\beta^\delta} \tensor{\phi}{_\gamma^\kappa} = 0 \,,
\end{align}
since in 4 dimensions the fully antisymmetric tensor $\delta^{\mu\nu\alpha\beta\gamma}_{\sigma\rho\lambda\delta\kappa}$ is zero (there are more indices than dimensions so two indices have to be repeated). Eq.~\eqref{eq:expr-L-Omega} then becomes
\begin{equation}
	L = -\frac{a_0}{8} \Omega_{1,0} + \frac{\Omega_{2,1}}{X^2} \qty(2 a_0 - \frac32 a_1) \,.
\end{equation}
One can see from \eqref{eq:def-Omegas} and \eqref{eq:def-GB-Lovelock} that $\Omega_{1,0} = 4 \mathcal{G}$. Therefore, by choosing $a_0 = -2$ and $a_1 = 4a_0 / 3  = -8/3$, one obtains \eqref{eq:G-total-derivative-Colleaux}.

We can now use the expression of $\mathcal{G}$ as a total derivative to express the action \eqref{eq:action-GB} as a Horndeski theory. Injecting this relation into Eq.~\eqref{eq:action-GB} and integrating by parts gives
\begin{equation}
	S_\text{GB}[g_{\mu\nu}, \phi] = \int \dd[4]{x} \sqrt{-g} \frac{2}{X} \dv{f}{\phi} \delta^{\mu\nu\alpha\beta}_{\sigma\rho\lambda\delta} \tensor{\phi}{_\alpha^\lambda} \phi_\beta \phi^\delta \qty(\tensor{R}{_\mu_\nu^\sigma^\rho} + \frac43 \frac{\tensor{\phi}{_\mu^\sigma} \tensor{\phi}{_\nu^\rho}}{X}) \,.
	\label{eq:action-GB-integparts}
\end{equation}
After expanding the products, one finds that the Lagrangian density $L_\mathcal{G}$ of \eqref{eq:action-GB-integparts} is
\begin{equation}
	L_\mathcal{G} = - \dv{f}{\phi}
	\begin{multlined}[t]
		\Big[ 8 R^{\mu\nu} \phi_{\mu\nu} + \frac{4}{X} \phi^\mu \phi_{\mu\nu} \phi^\nu - 4 R \Box\phi - \frac{16}{X} \tensor{R}{_\mu^\nu} \phi^\mu \phi^\rho \phi_{\nu\rho} - \frac{16}{3X} \tensor{\phi}{_\mu^\nu} \phi^{\mu\rho} \phi_{\rho\nu} \\
		+ \frac{8}{X} \Box\phi \phi_{\mu\nu}\phi^{\mu\nu}  + \frac{16}{X^2} \phi^\mu \phi^\nu \tensor{\phi}{_\mu^\rho} \tensor{\phi}{_\nu^\sigma} \phi_{\rho\sigma} + \frac{8}{X} R_{\mu\nu} \phi^\mu \phi^\nu \Box\phi \\
		- \frac{8}{3X} (\Box\phi)^3 - \frac{8}{X} R_{\mu\nu\rho\sigma} \phi^\mu \phi^\nu \phi^{\rho\sigma} - \frac{8}{X^2} \phi^\mu \phi_{\mu\nu} \phi^\nu \phi_{\rho\sigma} \phi^{\rho\sigma}  \\
		- \frac{16}{X^2} \phi^\mu \phi^\nu \tensor{\phi}{_\mu^\rho} \phi_{\rho\nu} \Box\phi + \frac{8}{X^2} \phi^\mu \phi_{\mu\nu} \phi^{\nu} (\Box\phi)^2 \Big]\,.
	\end{multlined}
\end{equation}
One can recognise several total derivatives:
\begin{equation}
	\nabla_\mu\qty(\frac{1}{X}) = -\frac{2}{X^2} \phi^\nu \phi_{\mu\nu} \qq{and} \nabla_\mu(\ln(X)) = \frac{2}{X} \phi^\nu \phi_{\mu\nu} \,.
\end{equation}
integrating by parts the terms containing these total derivatives and writing contractions of the Riemann tensors as commutators of derivatives, one obtains
\begin{equation}
	L_\mathcal{G} = \begin{multlined}[t]
		\dv{f}{\phi} \qty[-E_{\mu\nu} \phi^{\mu\nu} (8 + 4 \ln(X)) - \frac{4}{3X} (L_1^{(3)} - 3 L_2^{(3)} + 2 L_3^{(3)} ) ] \\
		+ \dv[2]{f}{\phi} \qty[2X \ln(X) R + 4 \ln(X) L_1^{(2)} + 4 (L_1^{(2)} - L_2^{(2)}) ] \\
		+ 2\dv[3]{f}{\phi} X (1 - 3 \ln(X)) \Box\phi - 2 \dv[4]{f}{\phi}  X^2 \ln(X) \,,
	\end{multlined}
	\label{eq:action-GB-developped}
\end{equation}
where the $L_i^{(j)}$ are the DHOST Lagrangians introduced in \cite{BenAchour:2016fzp}:  
\begin{align}
	&L_1^{(2)} = \phi_{\mu\nu} \phi^{\mu\nu} \,,\quad L_2^{(2)} = (\Box\phi)^2 \,,\nonumber\\
	&L_1^{(3)} = (\Box\phi)^3 \,,\quad L_2^{(3)} = (\Box\phi)\phi_{\mu\nu} \phi^{\mu\nu} \,,\quad L_ 3^{(3)} = \phi_{\mu\nu}\phi^{\nu\rho}\phi^\mu_\rho \,.
\end{align}
Finally, one can rewrite the term $E_{\mu\nu} \phi^{\mu\nu}$ using $\nabla^\mu E_{\mu\nu} = 0$ and writing contractions of the Ricci as commutators of derivatives, yielding
\begin{multline}
	\int\dd[4]{x} \sqrt{-g} E_{\mu\nu}\phi^{\mu\nu} \dv{f}{\phi} = \int\dd[4]{x} \sqrt{-g} \Big[\frac12 R \dv[2]{f}{\phi} + 2 (L_1^{(2)} - L_2^{(2)}) \dv[2]{f}{\phi} \\ - 3 X \Box\phi \dv[3]{f}{\phi} - 4X^2 \dv[4]{f}{\phi}\Big] \,.
	\label{eq:Einstein-phimunu-Horndeski}
\end{multline}
Putting Eq.~\eqref{eq:Einstein-phimunu-Horndeski} into Eq.~\eqref{eq:action-GB-developped}, one finally concludes that the action \eqref{eq:action-GB} is equivalent to a cubic Horndeski theory with
\begin{align}
	&G(\phi, X) = - 4 \dv{f}{\phi} \ln(X) \,,\quad F(\phi, X) = -2 X (2 - \ln(X)) \dv[2]{f}{\phi} \,,\nonumber\\
	&Q(\phi, X) = 2 X (7 - 3 \ln(X)) \dv[3]{f}{\phi} \,,\quad P(\phi, X) = 2 X^2 (3 - \ln(X)) \dv[4]{f}{\phi} \,.
\end{align}
This direct proof, which does not exist in the literature to the best of our knowledge,  complements the indirect proof given in \cite{Kobayashi:2011nu} based on the equivalence of  the equations of motion.

\section{Asymptotic behaviour near the horizon for the 4dEGB black hole}
\label{app:expr-xis}

In this Appendix, we give the explicit value of the coefficients $\zeta_1$ up to $\zeta_8$ appearing in \eqref{Ypolar4dGB}: 
\begin{align}
	\zeta_1 &= -\frac{4 i (2 \beta +1) \sqrt{-2 \beta ^2+\beta +1} \Omega }{4 (2 \beta  \Omega +\Omega )^2+(\beta -1)^2}\,,\nonumber\\
	\zeta_2 &= -\frac{4 i \sqrt{1-\beta }}{(2 \beta +1)^{3/2} \Omega \nu} \Big[
	\!\begin{multlined}[t]
		(\beta -1)^2 \left(6 \beta ^2-2 \beta -1\right) (2 \beta +1)^3 \Omega ^2 \\
		+(\beta -1)^4 \beta  (\beta  (2 \beta  (8 \lambda -1)+8 \lambda -5)+1) \\
		+4 (2 \beta (\beta +1)-1) (2 \beta +1)^5 \Omega ^4
		\Big]\,,
	\end{multlined}	
	\nonumber\\
	\zeta_3 &= -\frac{2 i (1-\beta )^{3/2} \beta }{\sqrt{2 \beta +1} (2 \beta  (\beta +1)-1) \Omega } \,,\nonumber\\
	\zeta_4 &= \frac{4 (1-\beta )^{3/2}}{\sqrt{2 \beta +1} \nu} \Big[
	\!\begin{multlined}[t]
		(\beta -1)^2 (2 \beta  (4 \beta  (\beta  (3 \lambda -1)+\lambda -2)+2 \lambda +1)+1) \\ 
		-4 (2 \beta  (4 \beta  (\beta  \lambda +\lambda +1)-2 \lambda +1)-1) (2	\beta  \Omega +\Omega )^2
		\Big]\,,
	\end{multlined}	\nonumber\\
	\zeta_5 &= \frac{4 \sqrt{1-\beta } \beta  \sqrt{2 \beta +1}}{2 \beta  (\beta +1)-1}\,,\nonumber\\
	\zeta_6 &= \frac{8 (\beta -1)^2 \beta }{4 (2 \beta +1)^3 \Omega ^2+(\beta -1)^2 (6 \beta +1)}\,,\nonumber\\
	\zeta_7 &= \frac{8 i \beta  \Omega  \left(12 (\beta +1) (2 \beta +1)^3 \Omega ^2+(\beta -1)^2 (\beta  (10 \beta +13)+7)\right)}{12 (\beta -1) (2 \beta +1)^4 \Omega ^2+3
		(\beta -1)^3 (6 \beta +1) (2 \beta +1)}\,,\nonumber\\
	\zeta_8 &= \frac{4 (\beta -1) \beta  \left(4 (2 \beta  \Omega +\Omega )^2+(\beta -1)^2\right)}{\sqrt{-2 \beta ^2+\beta +1} \left(4 (2 \beta +1)^3 \Omega ^2+(\beta -1)^2 (6
		\beta +1)\right)} \,,
\end{align}
with
\begin{equation}
	\nu = (2 \beta  (\beta +1)-1)  \left(4 (2 \beta  \Omega +\Omega )^2+(\beta -1)^2\right)^2 \,.
\end{equation}

\section{Asymptotical behaviour for the scalar field}
\label{app:asymp-behav-scalar}

In this Appendix, we study the linear perturbations of the scalar field about a fixed background. This corresponds to a \enquote{decoupling limit}, in which metric perturbations are zero and only the scalar field perturbations stay dynamical.

\subsection{Effective potential}

Let us consider the equation for the scalar field perturbation of the form
\begin{equation}
	c_2(r) \delta\varphi''(r) + c_1(r) \delta\varphi'(r) + c_0(r) \delta\varphi(r) = 0 \,.
	\label{eq:general-deltaphi-app}
\end{equation}
We aim to obtain the asymptotical behaviour of $\delta\varphi$ near some value $r_0$ of $r$, for example $r_0 = +\infty$. It is not possible to take directly the limit $r \longrightarrow r_0$ for each coefficient $c_i$, since one does not know in general how the first and second derivative of $\delta\varphi$ scale with respect to each other.

One therefore changes variables in order to obtain a simpler equation. Let us write
\begin{equation}
	\delta\varphi(r) = \kappa(r) \chi(r) \,,
	\label{eq:chgmt-delta-phi}
\end{equation}
and we have 
\begin{equation}
	c_2 \kappa \dv[2]{\chi}{r} + \qty(2\kappa'c_2 + \kappa c_1)\dv{\chi}{r} + \qty(\kappa'' c_2 + \kappa' c_1 + \kappa c_0) = 0 \,.
\end{equation}
One can then get rid of the first derivative by imposing
\begin{equation}
	\frac{\kappa'}{\kappa} = - \frac{c_1}{2c_2} \,.
	\label{eq:choice-kappa}
\end{equation}
The equation becomes
\begin{equation}
	\dv[2]{\chi}{r} + \qty(\frac{\kappa''}{\kappa} + \frac{c_1}{c_2} \frac{\kappa'}{\kappa} + \frac{c_0}{c_2}) \chi = 0 \,.
\end{equation}
Using \eqref{eq:choice-kappa} allows us to find the relation
\begin{equation}
	\frac{\kappa''}{\kappa} = \frac{c_2' c_1 - c_2 c_1'}{2 c_2^2} - \frac{c_1}{2c_2} \frac{\kappa'}{\kappa} = \frac{c_2' c_1 - c_2 c_1'}{2 c_2^2} + \qty(\frac{c_1}{2c_2})^2 \,,
\end{equation}
which finally leads to the equation,
\begin{equation}
	-\dv[2]{\chi}{r} + V_\chi(r) \chi = 0\qq{with} V_\chi(r) = \frac{ c_2 c_1' - c_2' c_1 }{2 c_2^2} + \qty(\frac{c_1}{2c_2})^2 - \frac{c_0}{c_2} \,.
\end{equation}
In order to obtain the behaviour near $r_0$, one can then decompose $V_\chi(r)$ around $r_0$ and solve directly for $\chi$. The solution will be the expansion of $\chi$ around $r_0$. One must then come back to $\delta\varphi$ by using Eqs.~\eqref{eq:choice-kappa} and~\eqref{eq:chgmt-delta-phi}.

\subsection{Application to the 4D Einstein-Gauss-Bonnet black hole}

The equation of motion for a scalar perturbation $\delta\varphi(t,r)$ with the metric perturbation fixed to zero is
\begin{equation}
	c_2(r) \pdv[2]{\delta\varphi}{r} + c_1(r) \pdv{\delta\varphi}{r} + 	c_0(r) \delta\varphi  = 0 \,,
\end{equation}
with
\begin{align}
	c_0(r) &= \frac{8\alpha\ell(\ell+1)}{r^2} \qty(4\sigma A + 2\sqrt{A}A -2r\sigma A' + \sqrt{A}(2-2rA' +r^2 A''))\,,\\
	c_1(r) &= 8\alpha\qty(4\sqrt{A} A' - r\sigma A'^2+ \sigma A(4A'-2rA''))\,,\\
	c_2(r) &= -16\alpha\sqrt{A}\qty(2A + 2\sigma A \sqrt{A} -r\sigma \sqrt{A}A')\,.
\end{align}
We observe that time does not appear in the equations and $\delta \varphi$ satisfies an elliptic equation rather than the expected 
hyperbolic equation. The fact that $\delta \varphi$ does not propagate could be related a strong coupling problem. 

Applying the reasoning presented in the previous section, one finds that the asymptotical behaviour of $\delta\varphi$ is
\begin{equation}
	\delta\varphi = A r^{-i\sqrt{1+\lambda}} + B r^{+i\sqrt{1+\lambda}} \,.
\end{equation}

We do not recover exactly the asymptotical behaviour found in \eqref{eq:asymb-scal-EGB} where both  metric and  scalar perturbations have been 
considered. However, the behaviours are very similar. 
This result  differs from the solutions studied in \cite{Langlois:2021aji}, for which the behaviour of the decoupled scalar perturbations and the scalar mode found from the full system agreed at both the horizon and infinity. It can be seen as the effect of a more important backreaction of the scalar field onto the metric. One can note that the behaviours still agree in the $\lambda \longrightarrow +\infty$ limit, implying that the coupling between the metric and the scalar perturbations becomes subdominant in that case.

\section{Equations of motion for the background and for axial perturbations}
\label{app:axial-first-order}

The variation of  the shift-symmetric Horndeski action \eqref{eq:generic-quintinc-Horndeski} yields the equations of motion 
\begin{equation}
	{\cal B}_{\mu\nu}\equiv \frac{\delta S}{\delta g_{\mu\nu}}=0 \,,\qquad  {\cal B}_{\phi}\equiv \frac{\delta S }{\delta \phi}=0 \, .
\end{equation}
Due to Bianchi identities, the equation for the scalar field is not independent from the metric equations and therefore can be ignored. For a metric of the form \eqref{metric} and a scalar field profile \eqref{eq:scal-ansatz}, one finds that there are only four non-trivial equations which are given in a supplementary Mathematica notebook.

Given any background  metric $\tensor{\overline{g}}{_\mu_\nu}$  solution to the above equations, one can introduce the perturbed metric 
\begin{equation}
	\tensor{g}{_\mu_\nu}=\tensor{\overline{g}}{_\mu_\nu}+ \tensor{h}{_\mu_\nu} \,,
\end{equation}
where the $\tensor{h}{_\mu_\nu}$ denotes the linear perturbations of the metric. In order to derive the linear equations of motion that govern the evolution of $h_{\mu\nu}$, one expands the action \eqref{eq:generic-quintinc-Horndeski} up to  second order in $ \tensor{h}{_\mu_\nu} $. The Euler-Lagrange equations associated with the \emph{quadratic} part $S_\text{quad}$ of this expansion then provide the linearised equations of motion for $h_{\mu\nu}$. In the following, they will be written under the form $\mathcal{E}_{\mu\nu} = 0$, where $\mathcal{E}_{\mu\nu}$ is defined by
\begin{equation}
	\mathcal{E}_{\mu\nu} = \fdv{S_\text{quad}}{h_{\mu\nu}} \,.
\end{equation}

In the Regge-Wheeler gauge, all the components of $h_{\mu\nu}$ for $\ell\geq 2$ are expressed in terms of the independent functions $h_0$ and $h_1$, as given in \eqref{eq:odd-pert}. In this gauge, one can show that the  equations of motion reduce to the  three equations $\mathcal{E}_{t\theta} = 0$, $\mathcal{E}_{r\theta} = 0$ and $\mathcal{E}_{\theta\theta} = 0$. These three equations depend only on $F$ and $G$, since the terms proportional to  $P$ and $Q$ and their derivatives vanish when the above background equations are imposed. They are given in the supplementary notebook as well.

As there are  only two independent functions, $h_0$ and $h_1$, one expects one of the above equations to be redundant. This is indeed verified by noting  the following relation between the equations and their derivatives:
\begin{equation}
	\dv{\mathcal{E}_{r\theta}}{r} - \frac{i\omega}{A B}  \mathcal{E}_{t\theta} + \frac{B'}{B} \mathcal{E}_{r\theta} + \frac{\lambda}{r^2 B} \mathcal{E}_{\theta\theta} = 0 \, .
\end{equation}
This shows that the two equations $\mathcal{E}_{r\theta}=0$ and $\mathcal{E}_{\theta\theta}=0$ are sufficient to fully  describe the dynamics of axial perturbations. Finally,  these two equations   can  be formulated as a simple first order system, given in \eqref{eq:system-axial-canonical}.

\bibliographystyle{utphys}
\bibliography{biblio_EGB}

\end{document}